\documentclass{article}
\usepackage{amssymb}
\usepackage{amsmath}
\usepackage{subfiles}
\usepackage{authblk} 
\usepackage{graphicx} 
\usepackage{capt-of}

\usepackage{array, booktabs} 
\newcolumntype{P}[1]{>{\centering\arraybackslash}p{#1}}

\DeclareMathOperator*{\SumInt}{%
\mathchoice%
  {\ooalign{$\displaystyle\sum$\cr\hidewidth$\displaystyle\int$\hidewidth\cr}}
  {\ooalign{\raisebox{.14\height}{\scalebox{.7}{$\textstyle\sum$}}\cr\hidewidth$\textstyle\int$\hidewidth\cr}}
  {\ooalign{\raisebox{.2\height}{\scalebox{.6}{$\scriptstyle\sum$}}\cr$\scriptstyle\int$\cr}}
  {\ooalign{\raisebox{.2\height}{\scalebox{.6}{$\scriptstyle\sum$}}\cr$\scriptstyle\int$\cr}}
}
\usepackage{caption}
\usepackage{subcaption}
\usepackage{pdflscape}
\usepackage{afterpage}
\usepackage{capt-of}
\usepackage{float}

\usepackage{hyperref}
\usepackage{xcolor} 
\usepackage{bookmark}

\usepackage{graphicx}

\newcommand{\txsub}[2]{ {#1}_{ \text{\tiny{#2} } } }

\newcommand{\uab}{^{ab}}
\newcommand{\dab}{_{ab}}

\newcommand{\bbR}{\mathbb{R}}

   \newcommand{\calD}{\mathcal{D}}                      

\newcommand{\sqrtdetg}{\sqrt{|g|}}

\DeclareMathOperator{\Tr}{Tr} 

\newcommand{\be}{\begin{eqnarray}}
\newcommand{\ee}{\end{eqnarray}}
\newcommand{\bea}{\begin{eqnarray}}
\newcommand{\eea}{\end{eqnarray}}

\def\del{\partial}

\title{On the One-Loop Exactness of Gravity Partition Function}

\author[a]{Andr\'es Goya \thanks{agoya@iafe.uba.ar}}
\affil[a]{Instituto de Astronom\'{\i}a y F\'{\i}sica del Espacio IAFE-CONICET, Ciudad Universitaria, IAFE, 1428, Buenos Aires, Argentina.}

\author[a, b]{Mauricio Leston \thanks{mauricio@iafe.uba.ar}}
\affil[b]{Departamento de F\'isica, Universidad de Buenos Aires FCEN-UBA and IFIBA-CONICET, Ciudad Universitaria, Pabell\'on 1, 1428, Buenos Aires, Argentina.}

\author[b]{Mario Passaglia \thanks{mpassaglia@cbc.uba.ar}}

\date{\today}

\begin{document}

\maketitle

\begin{abstract}
    In a previous work, we showed that the two- and three-loop contributions to the partition function of three-dimensional gravity in flat space vanish. This is in agreement with the expected one-loop exactness dictated by the underlying symmetry at the quantum level. To highlight the distinctive nature of the $D=3$ case, we extend the three-loop computation to arbitrary spacetime dimensions $D$. In higher dimensions, the number of contributions increases substantially, reinforcing the view that one-loop exactness is a unique feature of three-dimensional gravity.
\end{abstract}

\section{Introduction}
\label{sec:intro}

Einstein gravity in three dimensions has special features. One is the absence of local degrees of freedom. Another peculiarity appears at the quantum level: the partition function turns out to be one-loop exact.

In the case of gravity around AdS$_3$, the one-loop exactness was argued by Maloney and Witten in a paper in which they explicitly computed the sum over metrics in the Euclidean setting \cite{Maloney.Witten}. The argument is based on the analysis of the classical asymptotic symmetries in AdS$_3$ carried out by Brown and Henneaux \cite{Brown.Henneaux}. In summary, the idea is that the Hilbert space of a sector of quantum gravity organizes as Virasoro descendants; therefore, the partition function is a Virasoro character. From the full form of the character, one can read the effective action and conclude that the only correction to the classical action is of order zero in what plays the role of the Planck constant. This is precisely what one would obtain as the one-loop effective action in perturbation theory around AdS$_3$. This means that if one performs the perturbative computation of the partition function, it will terminate at one loop, with all the higher loop contributions vanishing. Although the status of pure gravity around AdS$_3$ as a consistent quantum theory remains under debate (see, for example, \cite{Cotler.Jensen:AdS3gravity.randomCFT} and \cite{MaloneyAverage, MaloneyConical}), the one-loop exactness rests on solid theoretical grounds.

For three-dimensional gravity in flat space, the idea that the partition function is one loop exact was first discussed by Witten in \cite{Witten:TopologyChaningAmplitudes}, where he performed a computation using the first-order formalism and argued that the perturbative series should terminate at one loop. More recently, in \cite{Barnich:One.Loop.Flat}, a similar argument to the AdS$_3$ case was presented for flat space, now relying on the asymptotic symmetry at null infinity, namely the Bondi-Metzner-Sachs (BMS) group. This line of reasoning suggests that the partition function should coincide with the BMS character. In analogy to the AdS$_3$ case, one can read the effective action and conclude that the only correction to the classical action is again of order zero in what plays the role of the Planck constant. Then, it follows that the partition function of three-dimensional Einstein gravity in flat space is indeed one-loop exact. As they pointed out, the one-loop exactness does not follow from the absence of local degrees of freedom, as one can see in theories with topological degrees of freedom, such as Chern-Simons theory \cite{Barnich:One.Loop.Flat}.

Despite the previous argument, the authors of \cite{Barnich:One.Loop.Flat} stated that it would be interesting to verify the one-loop exactness of three-dimensional gravity through a direct perturbative computation. We would like to understand how the phenomenon of one-loop exactness manifests in perturbation theory when one computes the vacuum diagrams. Is it the case that each individual diagram vanishes in $D=3$ due to specific features revealed through dimensional regularization? Or does the cancellation occur more subtly—perhaps through precise compensations between contributions from graviton interactions and those from ghost fields?

We carried out precisely this type of computation in a previous work \cite{Leston:ThreeLoops}, where we carefully worked out the Feynman diagrams up to three loops in the perturbative expansion for flat space gravity. We found that at two-loops, each diagram vanishes trivially in a generic dimension due to dimensional regularization. This result agrees with the literature \cite{Buchbinder:TwoLoopApproach, Buchbinder:TwoLoopsApproximation}. Therefore, up to two loops, nothing special occurs in $D=3$. However, a more complex contribution appears at three loops, given that several diagrams are not trivially zero and the number of terms associated with each diagram grows notably with the dimension. For simplicity, we started with the $D=3$ case in \cite{Leston:ThreeLoops}. We found that each non-vanishing diagram contributes with a given three-loop integral weighted with a coefficient leading to a beautiful zero:
\begin{figure}[h!]
\centering
\includegraphics[width=.9\textwidth]{Equal_to_zero_3D_one_line.pdf}
\end{figure}
\newline
The numbers below each diagram encode the total coefficient of a single three-loop integral, which includes the symmetry factor. The zero is a result of an intricate cancellation that takes place among the different diagrams appearing at three loops.

One might wonder whether this feature is exclusive to $D=3$ or not. It might seem that this result will also occur in generic dimensions, as it was the case at two loops, where the contribution was zero regardless of the dimension of the space-time. With that aim, we set out to compute the three-loop correction to the partition function around flat space in arbitrary dimension $D$.

The paper is organized in the following way: in section 2, we briefly review the main considerations regarding the partition function in the AdS$_3$/flat case. Then, in section 3, we write the action in a form suitable for our purposes. In section 4, we summarize the previous results of \cite{Leston:ThreeLoops} on the vanishing of the two- and three-loop computations in $D=3$ for the flat case. Finally, in section 5, we present the case of generic dimension. We conclude with some remarks.

\section{Brief review of the computation of the partition function of 3D gravity}

Before considering our specific result, we want to expand on the general argument given in the introduction concerning the computation and the group theoretical reasons leading to the conclusion that the partition function of 3D gravity around globally AdS space or flat space should be one-loop exact.

As in any quantum theory (in any spacetime dimension), in quantum gravity, the partition function
\begin{equation}\label{eq:Z}
    Z = \int \calD \,g\dab \, e^{-\txsub{S}{E}[g\dab]} \,,
\end{equation}
is a quantity of significant interest, given that it captures essential information about its spectrum. As we mentioned before, the three-dimensional model offers an appealing scenario in which certain computations can be performed more easily. In particular, for the negative cosmological constant case, Maloney and Witten \cite{Maloney.Witten} were able to compute the perturbative partition function in the AdS$_3$ sector by performing an explicit summation over classical configurations and their respective perturbations, whose boundary has the topology of a torus with $\tau$ as the modular parameter. If, for simplicity, we take $\tau = i\beta$, then the partition \eqref{eq:Z} will correspond to the integration of Euclidean geometries with period $\beta$, and from the Hilbert space perspective, \eqref{eq:Z} will coincide with the following trace:
\begin{equation}
    Z = \Tr e^{-\beta H} \,.
\end{equation}

In \cite{Maloney.Witten}, the computation explained above was done for generic complex $\tau$. For the particular sector of geometries connected with global AdS$_3$, a group theoretic argument can be made: the fact that the Hilbert space of this sector is comprised of Virasoro descendants of the vacuum (more technically, a coadjoint orbit of the Virasoro group) means that its partition function must be given by the vacuum Virasoro character. The result is
\begin{equation} \label{eq:vircharacter}
    \txsub{Z}{AdS}(\tau) = |q|^{-2k} \prod_{n=2}^{\infty} \dfrac{1}{|1-q^n|^2} \,,
\end{equation}
where $q=e^{2\pi i\tau}$ and $k \sim \frac{\ell}{G}$ ($\ell=$ ratio of AdS$_3$ and $G$ the Newton constant)\footnote{The full partition function is actually given by an infinite regularized sum over $\txsub{Z}{AdS}(\tau)$ and all the corresponding geometries related by the modular transformation group acting on $\tau$. The final result was not physically sensible, and this was explored, for instance, in \cite{MaloneyAverage, MaloneyConical}.}.

What is important to notice about \eqref{eq:vircharacter} is that its expression is 1-loop exact. This can be understood as follows: the effective action is given by $\txsub{I}{eff} = \txsub{I}{cl} + \sum_{n=1}^{\infty} k^{-n} I_n$ where $I_{cl}$ is the classical action and $I_n$ are the $n-$loop corrections. Simply by comparison with \eqref{eq:vircharacter} we realize that $I_{cl} = -4\pi k\, \rm{Im}{(\tau)}$ is the on-shell Einstein-Hilbert action\footnote{Actually, the on-shell Einstein-Hilbert action diverges, and it is necessary to add the Gibbons-Hawking term and an intrinsic boundary term to regularize it \cite{Balasubramanian.Kraus, HolographicRenormalization}.}, and the contribution of the vacuum descendants goes exactly like the 1-loop correction with no higher orders. Therefore, the contribution to the Euclidean partition function coming from the AdS$_3$ orbit is one-loop exact (see also \cite{Cotler.Jensen:reparametrization.AdS3}, where the authors derive the one loop exactness from localization arguments). The one-loop contribution \eqref{eq:vircharacter} was computed explicitly as the appropriate functional determinant in \cite{Giombi:One.Loop} and the result agrees with the one expected from the Virasoro character. In connection with that, the survival of the Virasoro symmetry after the gauge fixing procedure was clarified in \cite{Acosta:SquareIntegrability} and its accessibility in \cite{Acosta:Accesibility}.

An analogous computation can be performed in the case of zero cosmological constant. In \cite{Barnich:One.Loop.Flat} the flat space one-loop partition function was computed, taking advantage of the fact that the states of the gravity theory organize into representations of the asymptotic symmetry group; in this case, is the BMS$_3$ group \cite{Ashtekar:BMS, Barnich:BMS}. A different perspective on the matter can be found in \cite{Cotler.etal:soft.gravitons.3d}. From the BMS character, as in the case of AdS$_3$, it can be inferred that the partition function is one-loop exact.

\section{Perturbative expansion in flat gravity}
In this section, we review the perturbative expansion of the Einstein-Hilbert action in a manner suitable for our purposes.

The gravitational action is given by
\begin{equation}\label{eq:SEH}
    \txsub{S}{EH} = -\dfrac{1}{2\kappa^2} \int{d^dx  \sqrtdetg \, \left( R - 2\Lambda \right)} + \txsub{S}{bdy}\,,
\end{equation}
where $\txsub{S}{bdy}$ is an appropriate boundary term and $\kappa^2=8\pi G$, with $G$ the Newton constant.

It turns out that the following is a very convenient way to rewrite the Einstein-Hilbert action \eqref{eq:SEH}
\begin{equation}\label{eq:SEHLandau}
\begin{split}
    \txsub{S}{EH} = -\frac{1}{2\kappa^2}&\int_{M_d} d^d x \sqrtdetg \, g^{ab} g^{mn}  g^{rs} \Big(  \partial_{m}{g_{ab}} \partial_{n}{g_{rs}} -\\ & 
     \partial_{m}{g_{ar}} \partial_{n}{g_{bs}}  + 2 \partial_{m}{g_{br}} \partial_{a}{g_{ns}} - 2 \partial_{m}{g_{na}} \partial_{b}{g_{rs}}   \Big) \,,
\end{split}
\end{equation}
up to boundary terms \cite{Landau.Lifschits:ClassicalFields}. 

Next, we need to expand the action around a spacetime background that, in our case, is flat space
\begin{equation}
    g\dab = \eta\dab + \kappa h\dab \,,
\end{equation}
where $\eta$ is the Minkowski metric. Replacing this into \eqref{eq:SEHLandau} we obtain
\begin{equation}
\begin{split}
\label{eq:EHactionLandauh}
    \txsub{S}{EH} =&-\frac 12  \int_{M_{d}} d^d x \sqrt{|g|} \, g^{ab} g^{mn} g^{rs} \Big( \partial_{m}{h_{ab}} \partial_{n}{h_{rs}} -\\ &  \partial_{m}{h_{ar}} \partial_{n}{h_{bs}}  + 2 \partial_{m}{h_{br}} \partial_{a}{h_{ns}} - 2 \partial_{m}{h_{na}} \partial_{b}{h_{rs}}   \Big) \,.
\end{split}
\end{equation}
What is particularly useful about the above expression is the following: to obtain any graviton vertex at any perturbation order, we only need to expand the inverse metric $g\uab$ and the measure $\sqrtdetg$ in terms of $h\dab$. In particular, the graviton propagator turns out to be
\begin{equation}
\label{eq:gravitonprop}
    \Delta_{\rm (gr)}^{mnab}[k] = \dfrac{i}{2k^2} \left( \eta^{ma}\eta^{nb} + \eta^{mb}\eta^{na} - \dfrac{2}{d-2}\eta^{mn}\eta^{ab} \right)\, 
\end{equation}
with $k$ being the momentum.

As usual, it is necessary to introduce a gauge-fixing term
\begin{equation}
\label{gaugefix}
    \txsub{S}{gf} = \int_{M_d} d^dx \, f^{m} \eta_{mn} f^{n} \,,
\end{equation}
where the gauge-fixing function is
\begin{equation}
\label{harmonicgauge}
    f^{m} = \left( \eta^{lm} \bar{\del}^{n} - \frac12 \eta^{ln} \bar{\del}^{m} \right) h_{nl} \, ;
\end{equation}
and the corresponding ghost field action is
\begin{equation}
\label{eq:ghostaction}
    \txsub{S}{gh} =  \int_{M_d} d^d x \, \bar{c}^{m} \,\dfrac{\delta f_{m}}{\delta h_{rs}} \, \mathcal{L}_{c} g_{rs} \,,
\end{equation}
where $c^s$ and $\bar{c}^l$ are the ghost and anti-ghost fields, with ${\mathcal L}_{c} g_{rs}$ being the Lie derivative of the full metric $g_{ab}$ with respect to the ghost field $c^l$
\begin{equation}
    \label{Liedrivghost}
    {\mathcal L}_{c} g_{rs} = 2 \eta_{l(s}\del_{r)}{c^l} + c^l\del_{l}{h_{rs}} + 2 h_{l(s}\del_{r)}{c^l} \,.
\end{equation}

More explicitly, up to a total derivative, the ghost action takes the form
\begin{equation}
\label{eq:ghostlagrangiannexp}
\begin{split}
    \txsub{S}{gh} = 
      \int_{M_d} d^d x \,
    \Big[  &\eta_{ls} \Bar{c}^{s} \del^2{c^l} + \bar{c}^{r}\del_{l}\del_{r}{c^l} - \bar{c}^{m}\del_{m}\del_{l}{c^l}  \\ 
    &- \kappa \Big(  \del^{r}{\bar{c}^{s}} \del_{l}{h_{sr}} c^l + \del^{r}{\bar{c}^{s}} h_{ls} \del_r{c^l}   
    +\del^{s}{\bar{c}^{r}} h_{ls} \del_{r}{c^l} \\
    & - \frac12 \del_{m}{\bar{c}^m}\del_{l}{{h}_{r}^{\, r}}\,c^l - \del_{m}{\bar{c}^{m}} h_{ls} \del^{s}{c^l} \Big)\Big] \,.
\end{split}
\end{equation}
The corresponding ghost propagator is
\begin{equation}
\label{eq:ghostprop}
    \Delta_{\rm (gh)}^{ab}[k] = -i \dfrac{\eta^{ab}}{k^2}\, .
\end{equation}
Again, $k$ is the momentum.

\subsection*{Remarks}
\begin{figure}[h]
\centering
\includegraphics[width=0.95\textwidth]{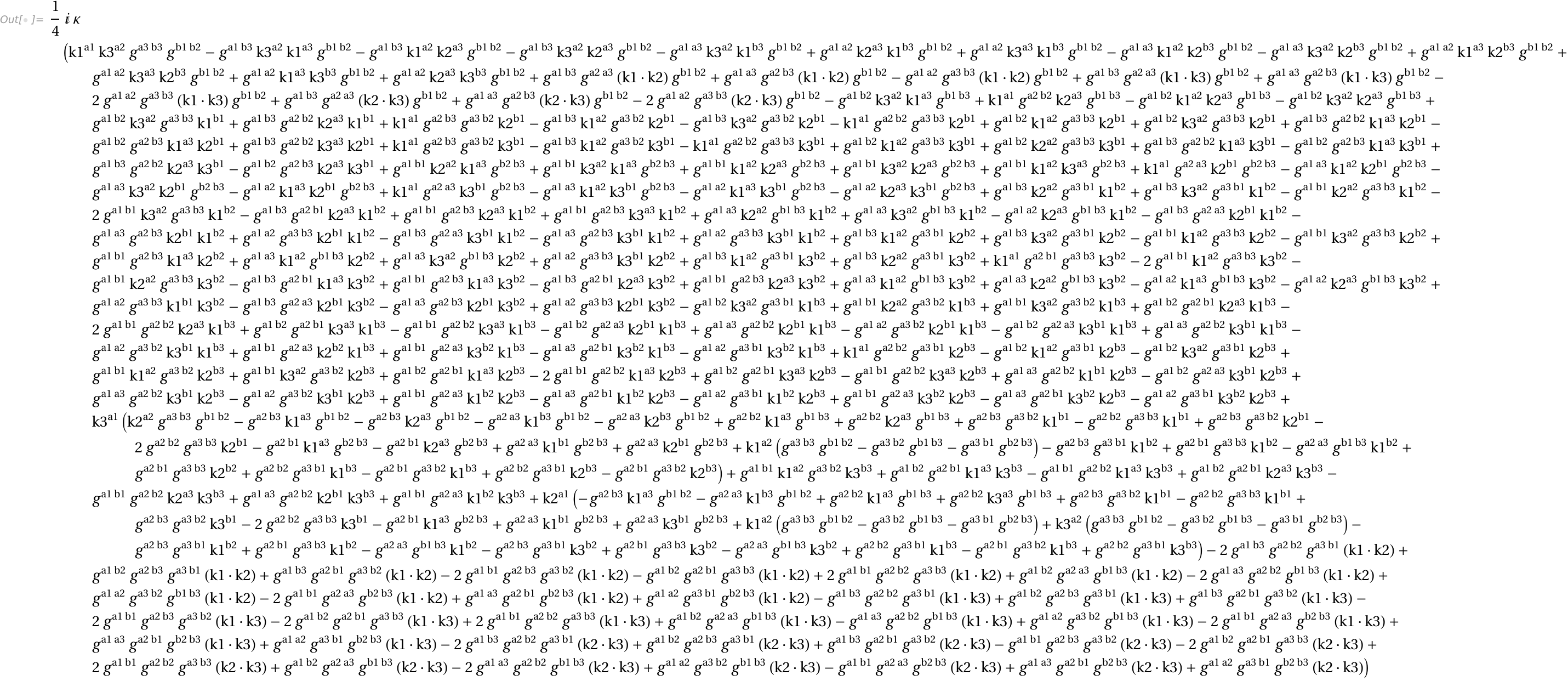}
\caption{Completely symmetrized expression of the three-graviton vertex.}
\label{fig:3gravitonvertex}
\end{figure}

What is important to notice is that:
\begin{itemize}
    \item all the vertices, both graviton and ghost, are quadratic in the momenta,
    \item there is a single ghost vertex which involves a ghost, a anti-ghost and one graviton leg.
\end{itemize}
The expression of the graviton vertex increases in complexity with the number of legs. Just for illustration purposes, we show the three-graviton vertex (including symmetrization terms) for momenta $k_1$, $k_2$ and $k_3$ in Fig. \eqref{fig:3gravitonvertex}.

In the three-loop computation, it is necessary to consider up to the five graviton vertex\footnote{Fortunately, the only diagram appearing at three loop level with a six graviton vertex is trivially zero by dimensional regularization.}. This explains why the computation is so lengthy.

\section{The partition function at three loops in \texorpdfstring{$D=3$}{D=3}}
Now, we briefly go over our work \cite{Leston:ThreeLoops}. We should notice that the computation is performed in the Euclidean signature; that is, we will consider as background $\bbR^3$ with the time coordinate identified with period $\beta$. We will also perform a Wick rotation in the timelike component of the momentum $k^{a}=(k^0\,,k^1\,,k^2) \rightarrow (k^3\,,k^1\,,k^2)$ where $k^3 = -ik^0$. The direct consequence of this is that the Euclidean \textit{timelike} component of the momentum is quantized as $k^3_{(n)} = \frac{2\pi n}{\beta}$. Therefore, the integrals $\int d^3k$ are actually composed of a sum over the Matsubara frequency and a 2-dimensional integral. For that reason, sometimes the symbol $\SumInt$ is used:
\begin{equation}
    \SumInt d^3k \equiv \beta\sum_{n} \int d^2k \,,
\end{equation}
or in $D$ dimensions:  $\SumInt d^Dk \equiv \beta\sum_{n} \int d^{D-1}k$

The partition function $Z$ is related to the effective action $\txsub{S}{eff}$ through
\begin{equation}
    -\log Z = \frac{1}{\kappa^2} \txsub{S}{eff} \,,
\end{equation}
with $\kappa^2$, playing the role of the $\hbar$, controlling the perturbative expansion
\begin{equation}
    -\log Z = \sum_{n=0}^{\infty} \kappa^{2(n-1)}\txsub{S}{eff}^{(n)} \,.
\end{equation}
$\txsub{S}{eff}^{(0)}$ is the on-shell action, $\txsub{S}{eff}^{(1)}$ corresponds to the 1-loop determinant, $\txsub{S}{eff}^{(2)}$ and $\txsub{S}{eff}^{(3)}$ are the 2- and 3-loop contributions respectively. 

\begin{figure}
\centering
\includegraphics[width=60mm]{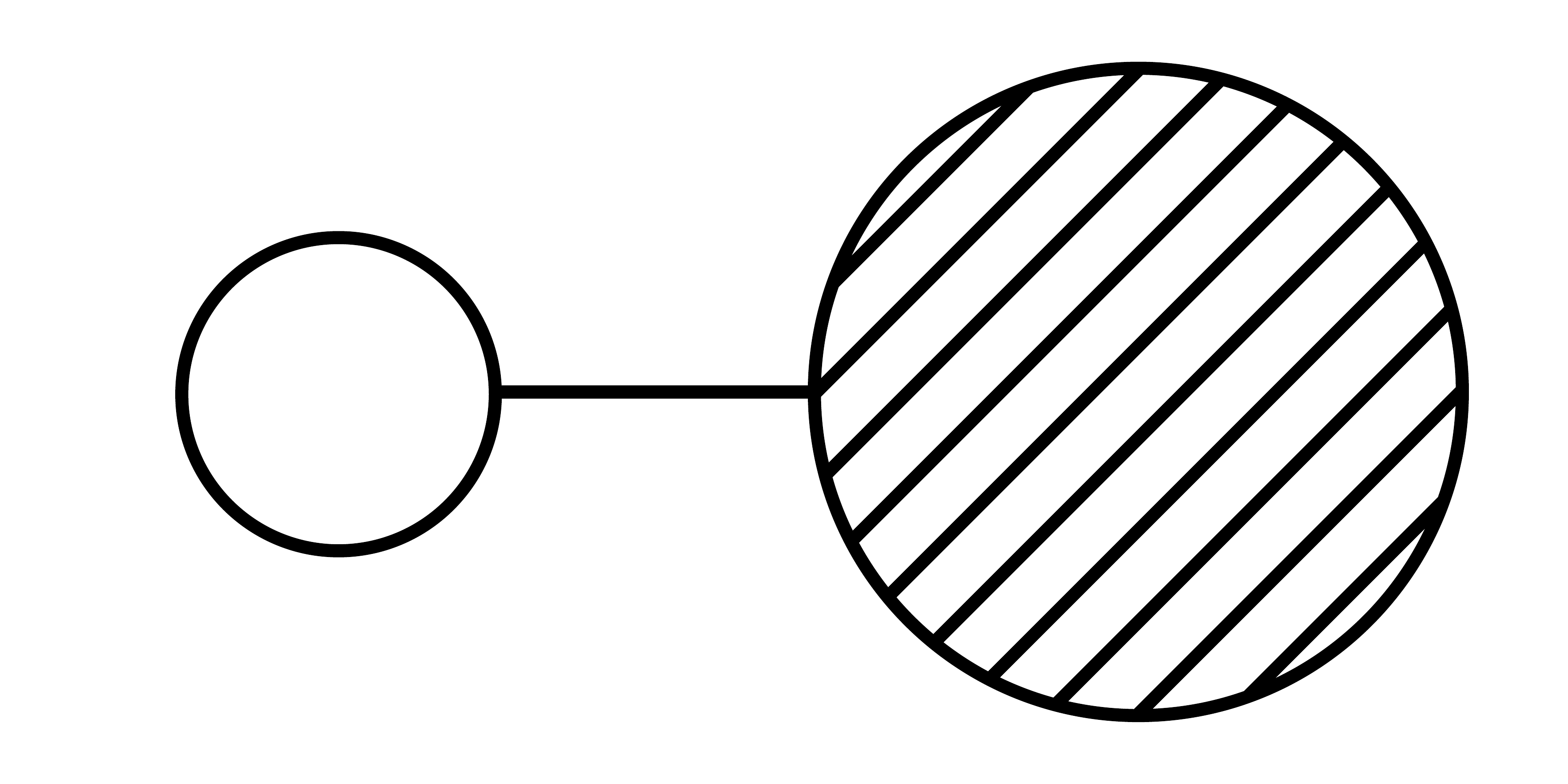}
\caption{Tadpole-like diagram vanishing by dimensional regularization. The same remark apply by replacing the graviton loop by a ghost propagator loop.}
\label{tadpole}
\end{figure}

In order to compute the effective action we need to consider all the connected vacuum diagrams, including those that are not 1PI. However, up to three-loops, all non 1PI diagrams have a tadpole-like subgraph, as depicted in the Figure \eqref{tadpole}.

These type of diagrams are zero by dimensional regularization. It can be easily seen by recalling that all the vertices (graviton and ghost) are quadratic in the momenta and therefore the contribution of this loop will be given by the following integral/sum:
$$\SumInt d^Dp \frac{p^2}{p^2}=\SumInt d^Dp= \beta\sum \int d^{D-1}p \,,$$
which vanishes by the {\it Veltman lemma} $\int dk\; k^n=0$, a key result of dimensional regularization that will be used use frequently in this work.

\subsection{The boring vanishing of the two loops contribution}
\begin{figure}[h!]
     \centering
     \begin{subfigure}[b]{0.3\textwidth}
         \centering
         \includegraphics[width=\textwidth]{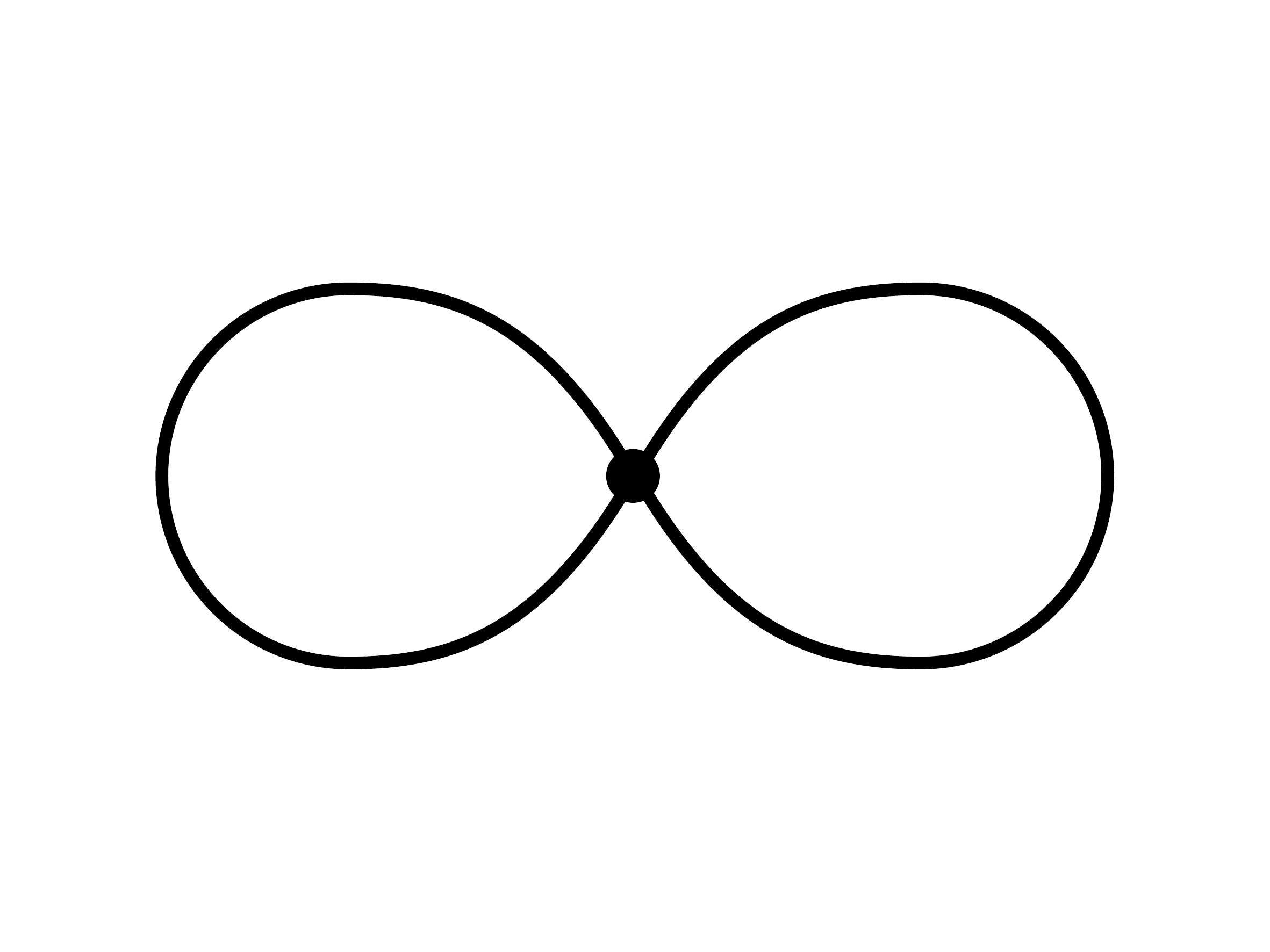}
         \caption{Infinity}
         \label{fig:twoloopinfinity}
     \end{subfigure}
     \hfill
     \begin{subfigure}[b]{0.3\textwidth}
         \centering
         \includegraphics[width=\textwidth]{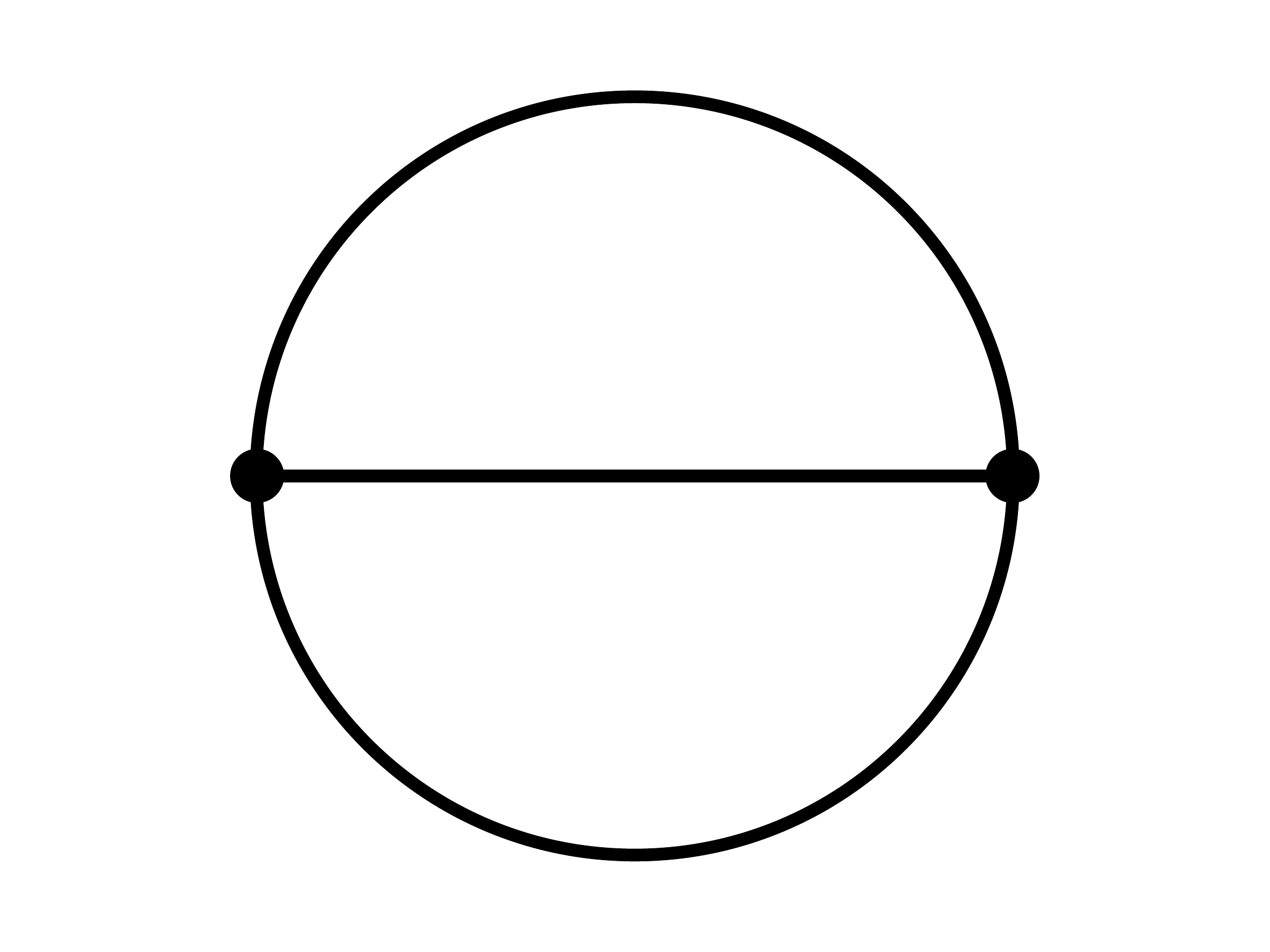}
         \caption{Sunset}
         \label{fig:twolooppeach}
     \end{subfigure}
     \hfill
     \begin{subfigure}[b]{0.3\textwidth}
         \centering
         \includegraphics[width=\textwidth]{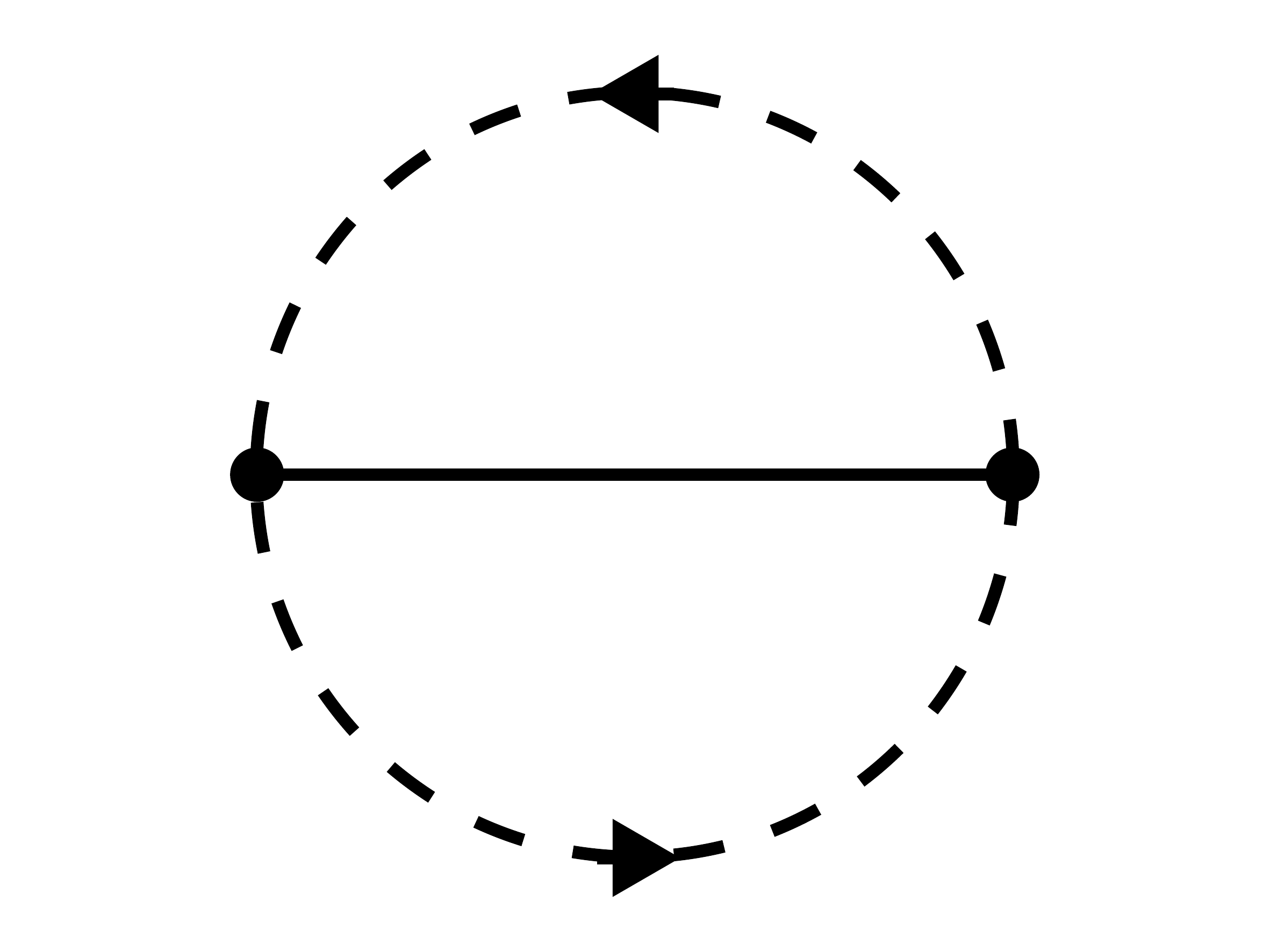}
         \caption{Ghostly sunset}
         \label{fig:twoloopspeachghost}
     \end{subfigure}
        \caption{Two-loop diagrams.}
        \label{fig:twolooppeachghost}
\label{fig:twoloops}
\end{figure}

When expanding the effective action, the two-loop contribution comes from the diagrams shown in Figure \eqref{fig:twoloops}.
One can see that these contributions trivially vanish not only in $D=3$ but in any dimension, by noticing that one integral always reduces to $\int dk \, 1$ or turns out to be odd in the momentum. For instance: 

\begin{equation}
    \begin{split}
    \SumInt \prod_{i=1}^2d^Dk_i\frac{(k_1 \cdot k_2)^2}{k_1^2k_2^2(k_1+k_2)^2} &=\SumInt \prod_{i=1}^2d^Dk_i  \frac{1}{4}\frac{\left((k_1 + k_2)^2 -k_1^2-k_2^2\right)^2}{k_1^2k_2^2(k_1+k_2)^2} \\
    &=\frac{1}{4}\SumInt d^Dk_1d^Dk_2 \left( \frac{1}{k_1^2} +\frac{1}{k_2^2} + \ldots \right) = 0 \,,
    \end{split}
\end{equation}
where the dots refer to similar terms containing the integral of $1$ in $k_1$ or $k_2$ (vanishing by the Veltman lemma) or either $\frac{k_1\cdot k_2}{k_1^2k_2^2}$ (vanishing by symmetry).

Therefore:
\begin{equation}
    \txsub{S}{eff}^{(2)} = 0 \,,
\end{equation}
in arbitrary dimension.

\subsection{Three-loop contribution in \texorpdfstring{$D=3$}{D=3}}
The three-loop contribution to the partition function is significantly more complex. This can be seen from the variety of diagrams that appear at this perturbation order (refer to Figs. \eqref{fig:threeloopnoghost}, \eqref{fig:threeloopsnowman}, \eqref{fig:threeloopcylinder}, and \eqref{fig:threeloopMB}. Although some diagrams turn out to be zero under dimensional regularization (for example, the diagrams \eqref{fig:trefoil}, \eqref{fig:turtle}, \eqref{fig:bear}, \eqref{fig:snowman}, and \eqref{fig:snowmanghost}), the remaining graphs contribute with several terms that are non-zero three-loop integrals. After lengthy work, discarding integrals that are zero by dimensional regularization, one can reduce all of them to a single master integral $I$:
\begin{equation}
\label{eq:I.integral.2}
        I = \SumInt \prod_{i=1}^3d^3 k_i \frac{(k_1 \cdot k_2)(k_1 \cdot k_3)}{k_1^2k_2^2k_3^2(k_1+k_2+k_3)^2} \,.
\end{equation}

So all non-zero diagrams contribute to the partition function with a term proportional to \eqref{eq:I.integral.2}. This is summarized in table \eqref{tab:diagrams.contribution.3d}. The three-loop effective action is
\begin{table}[H]
    \centering
        \begin{tabular}{m{1em} m{1em} m{0.8\textwidth} } 
            $\txsub{S}{eff}^{(3)}$ & $=$ & \includegraphics[width=.7\textwidth]{Equal_to_zero_3D.pdf} \\
          & $=$ & $\left( - \frac{61}{16} + 1 - \frac{1}{2} - \frac{7}{4} - \frac{3}{8} + \frac{3}{2} + \frac{11}{4} - \frac{13}{8} + \frac{45}{16} \right) I \quad = 0 \quad \,.$
        \end{tabular}
\end{table}

As we anticipated, the sum of the diagrams yields a vanishing result in $D=3$.

\section{The partition function in \texorpdfstring{$D$}{D} dimensions}
The argument given by \cite{Maloney.Witten, Barnich:One.Loop.Flat} leading to the one-loop exactness of the partition function used specific features of the algebraic structure of the three dimensional case. We want to determine whether the vanishing of the three-loop contribution to the effective action is something specific to the $D=3$ case. 

The generic $D$ dimensional case is much more complex. Several diagrams that vanished in the $D=3$ case no longer do so, with the notable exception of the trefoil diagram \eqref{fig:trefoil}, which remains zero in any dimension. More importantly, the diagrams now contain many additional terms and do not simplify to a single master integral, as they did in three dimensions. Instead, all momentum integrals decompose into linear combinations of several master integrals: $I\,, J_1\,, K_1\,, K_2\,, A\,, B\,,C$ \eqref{eq:master.integrals}. This result was obtained after a lengthy and tedious computation. The individual contributions of each diagram are summarized in Tables \eqref{tab:diagrams.contribution.I} and \eqref{tab:diagrams.contribution.D}.

Collecting all the terms, the three-loop effective action is given by
\begin{equation}\label{eq:S3loopsD}
    \begin{split}
        \rm{S_{eff}^{(3)}} &= (D-3) \Bigg{[} \frac{D^2 (D+6) (2 D-9)}{128 (D-2)}I + \frac{D \left(18 D^3+80 D^2-518 D-107\right)}{576 (D-2)}J_1 \\
        &- \frac{D \left(18 D^3 -126 D^2 +71 D +362\right)}{576 (D-2)}K_2 - \frac{(D-3) D^2 \left(D^2-4 D+2\right)}{64 (D-2)}K_1 \\
        &+ \frac{D^2 \left(D^2-3 D-6\right)}{64 (D-2)}A - \frac{(D-6)D^2}{256 (D-2)}B  - \frac{(D-3) D^2}{64}C \Bigg{]} \,.
    \end{split}
\end{equation}

One can see that each coefficient is a polynomial in $D$ which has a natural root only in $D=3$ (except for the one accompanying $B$). Of course, without knowing the precise relations among the integrals $I\,, J\,, \ldots\,, C$ (depending on $D$) one cannot claim definitively that this contribution vanishes only in $D=3$. A more detailed analysis of these three-loop integrals, which are generically divergent, is required. Nonetheless, this result indicates that the special nature of the partition function computation lies in the appearance of dimension-dependent polynomials that vanish specifically at $D=3$. This suggests that there is an underlying algebraic mechanism unique to three dimensions that enables these remarkable cancellations.

\section{Concluding remarks}
One of the main challenges in the computation lies in the huge number of terms involved. Each graviton vertex contains multiple contributions, all of which are quadratic in the momenta. As previously mentioned, the three-graviton vertex—depicted in Figure \eqref{fig:3gravitonvertex}—is the simplest case. This gives an idea of how tedious the calculation becomes for more complex diagrams, such as the Mercedes-Benz diagram \eqref{fig:MB}, which involves four such vertices. Even more intricate is the Glasses diagram \eqref{fig:glasses}, which includes a four-leg vertex that barely fits on the page.

A second major difficulty arises in reducing the resulting expressions to a set of master integrals, discarding those that vanish under dimensional regularization. This is where the number of dimensions becomes crucial: in $D=3$, many terms are absent. This simplification was precisely the reason why our earlier work \cite{Leston:ThreeLoops} began with the case $D=3$.

Our ultimate goal is to understand the mechanism behind the cancellation of contributions in three dimensions. While the algebraic structure of the Hilbert space already explains why the partition function must be one-loop exact, it remains an open question how this exactness emerges diagrammatically. In the previous work, we took a small step towards explicitly verifying the vanishing of the three-loop contribution in $D=3$. Here, we encountered significantly more involved contributions appearing in $D>3$ \eqref{eq:S3loopsD}. The calculation in $D=3$ reveals a nontrivial and intricate cancellation among diagrams, offering no immediate intuition for how this occurs. Understanding the deeper diagrammatic structure responsible for these cancellations is an interesting direction that we leave for future work.

\section*{Acknowledgements:} The authors thanks K. Jensen for useful comments. The computation presented in this paper partially resorted to FeynCalc \cite{FeynCalc1, FeynCalc2, FeynCalc3}. The computational resources used in this work were provided in part by the HPC center DIRAC, funded by IFIBA (UBA-CONICET) and part of SNCAD-MinCyT initiative, Argentina.


\appendix
\section{Vacuum diagrams at three loops}\label{diagramasvacio}
\begin{figure}[h!]
\centering
    \begin{subfigure}[b]{0.4\textwidth}
        \includegraphics[width=\textwidth]{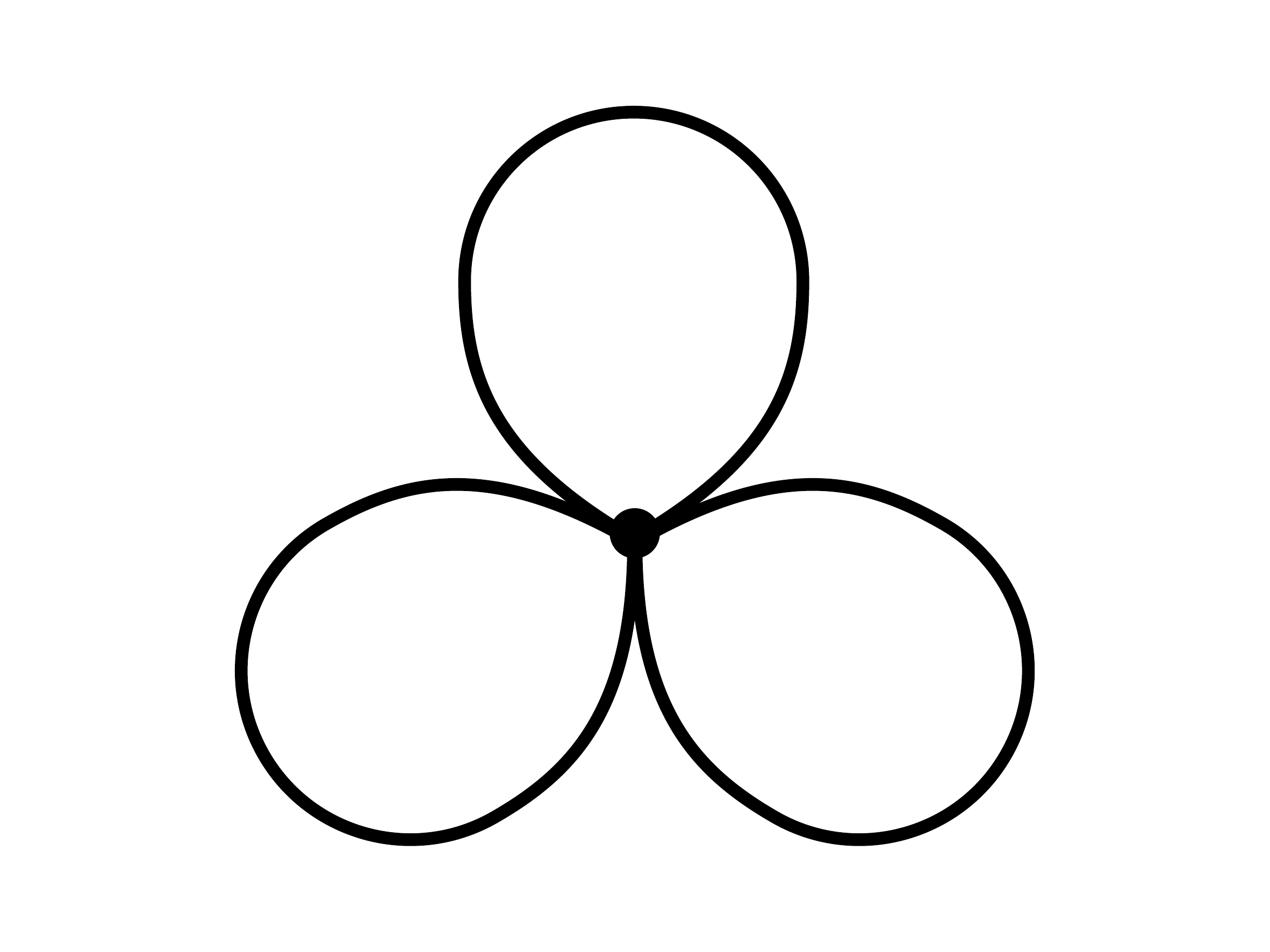}
        \caption{Trefoil}
        \label{fig:trefoil}
    \end{subfigure}
    \hfill
    \begin{subfigure}[b]{0.4\textwidth}
        \includegraphics[width=\textwidth]{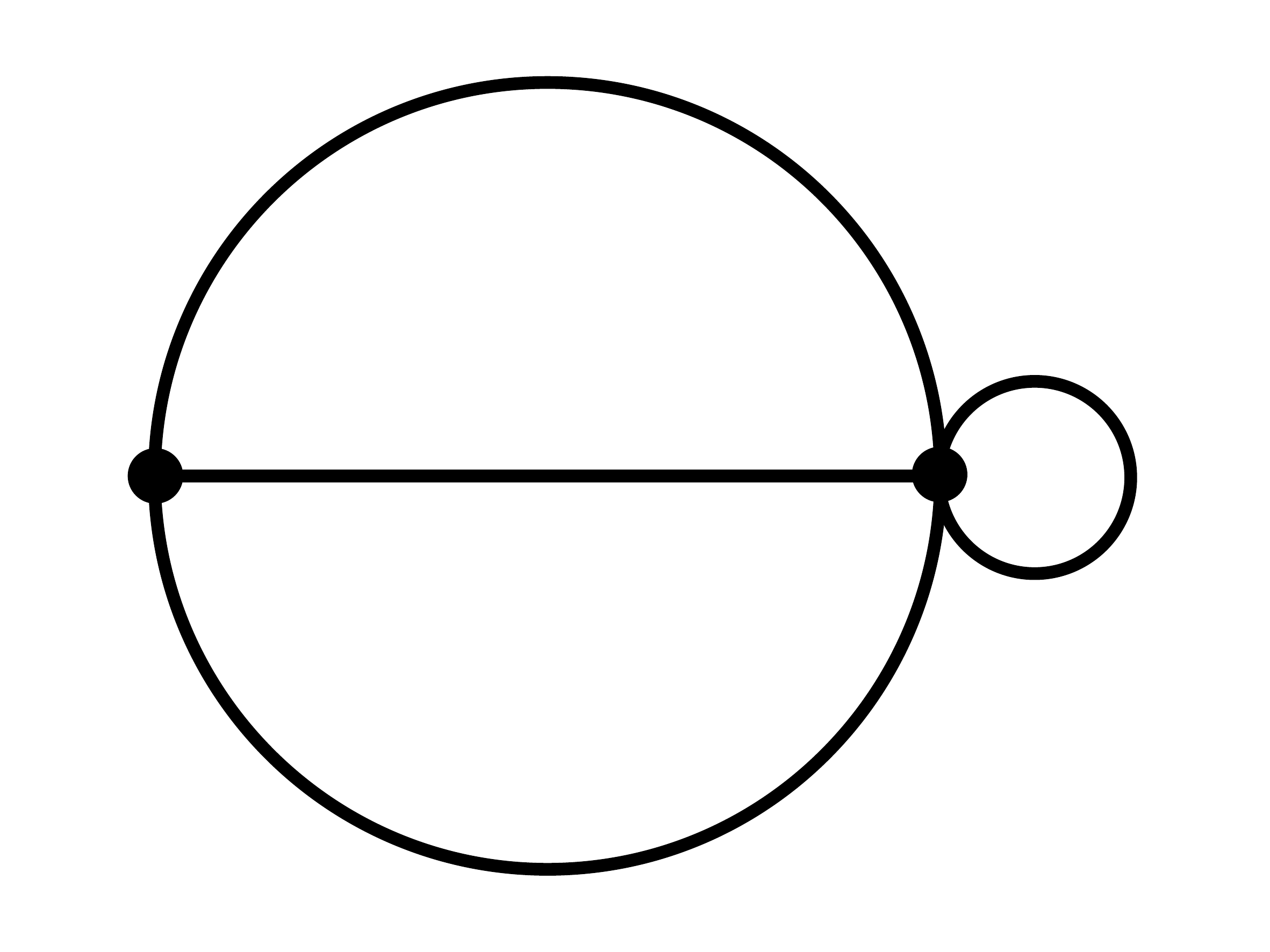}
        \caption{Turtle}
        \label{fig:turtle}
    \end{subfigure}
    \begin{subfigure}[b]{0.4\textwidth}
        \includegraphics[width=\textwidth]{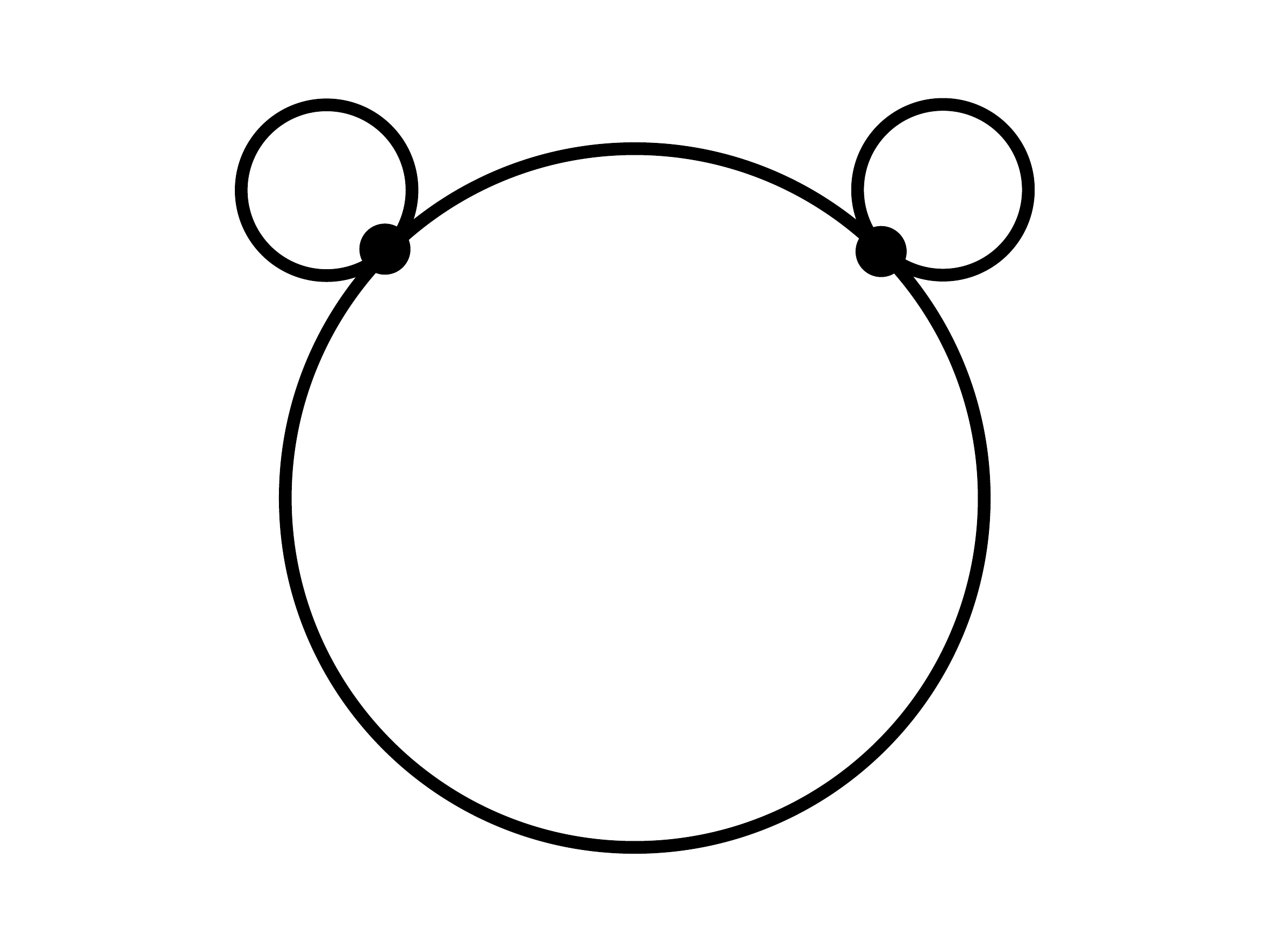}
        \caption{Bear}
        \label{fig:bear}
    \end{subfigure}
    \hfill
    \begin{subfigure}[b]{0.4\textwidth}
    \centering
        \includegraphics[width=\textwidth]{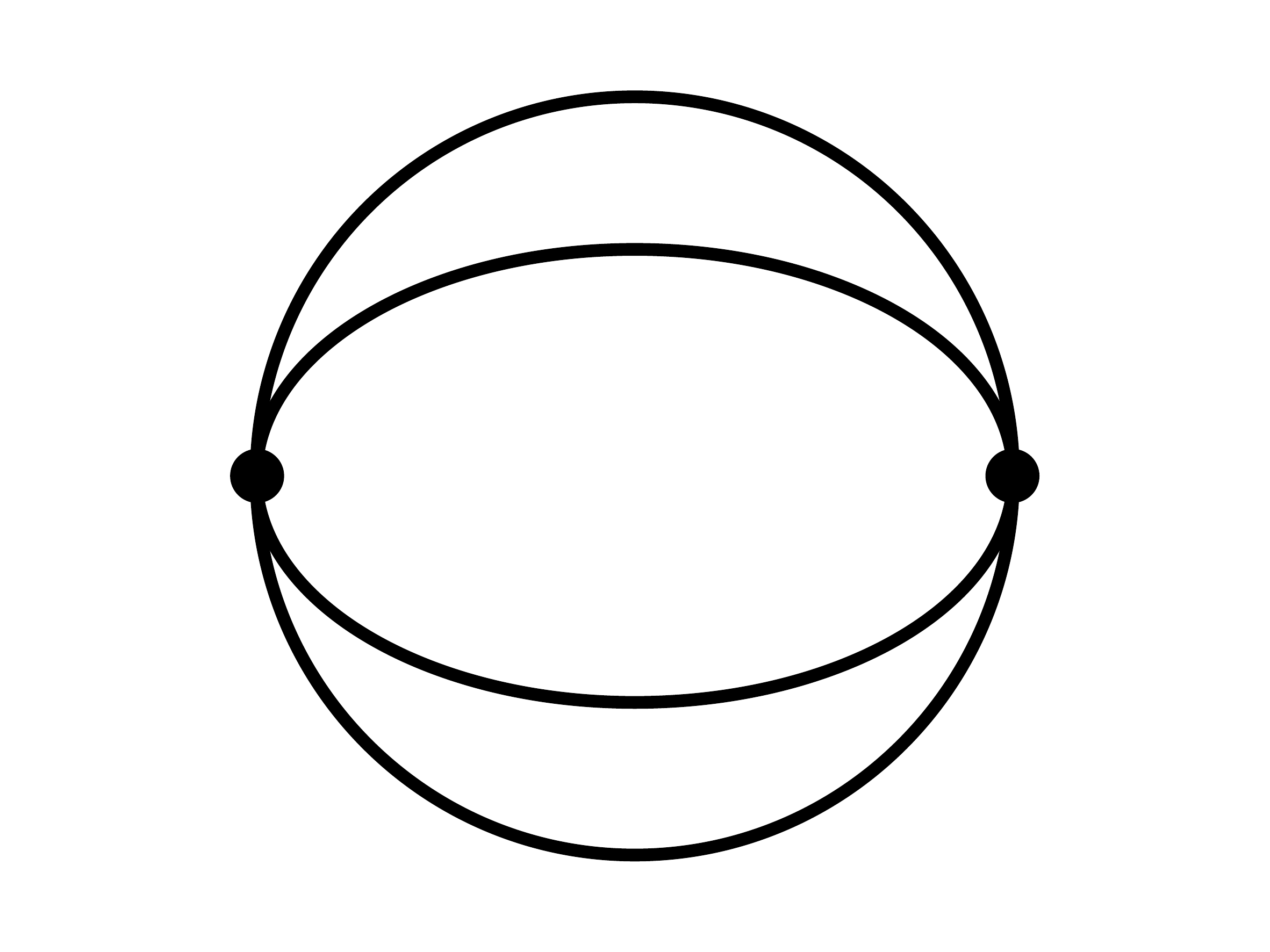}
        \caption{Mellon}
    \end{subfigure}
    \hfill
    \begin{subfigure}[b]{0.4\textwidth}
    \centering
        \includegraphics[width=\textwidth]{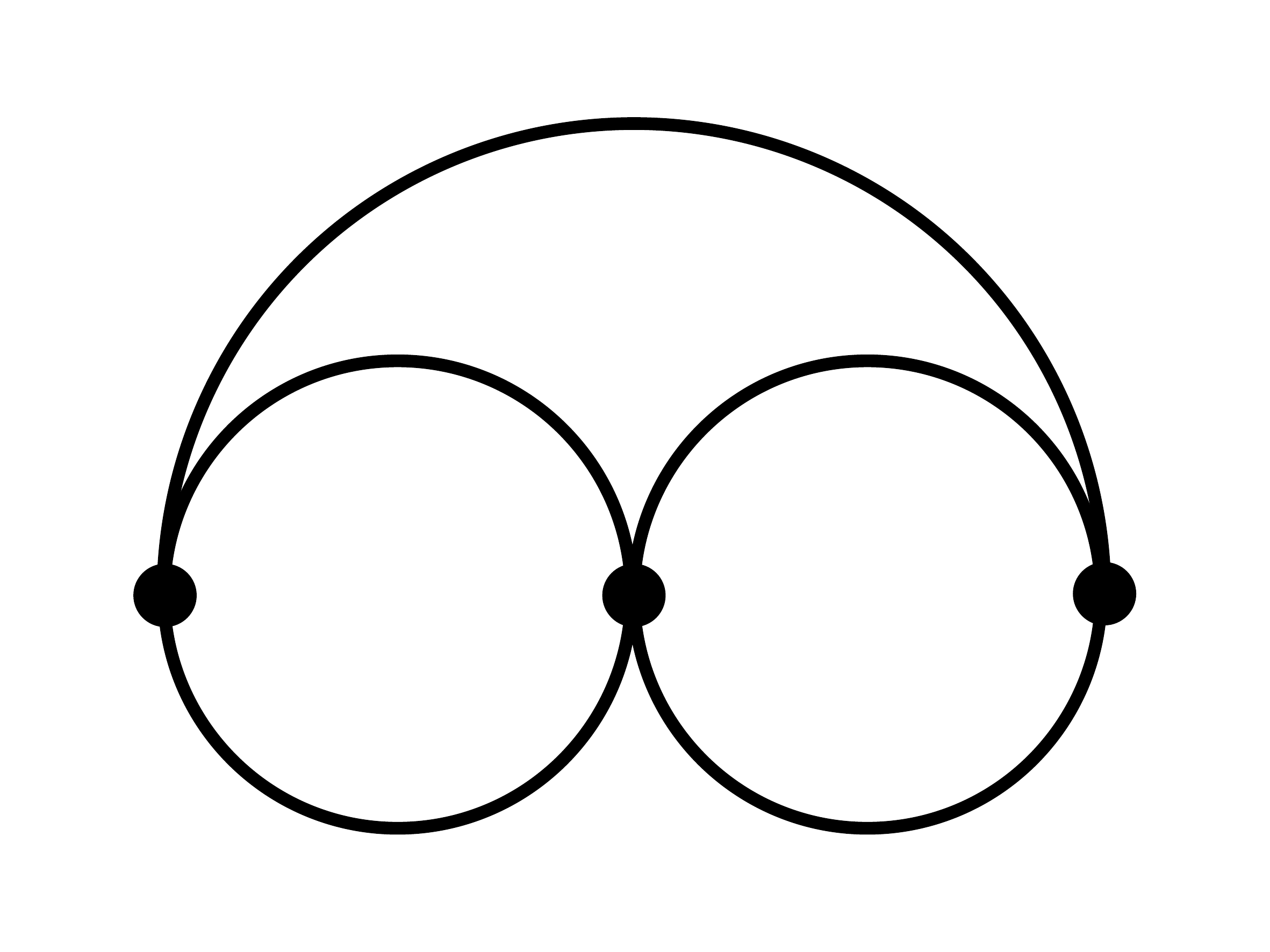}
        \caption{Glasses}
        \label{fig:glasses}
    \end{subfigure}
    \caption{Three-loop contributions}
\label{fig:threeloopnoghost}
\end{figure}

\begin{figure}[ht]
    \centering
    \begin{subfigure}[b]{0.4\textwidth}
        \includegraphics[width=\textwidth]{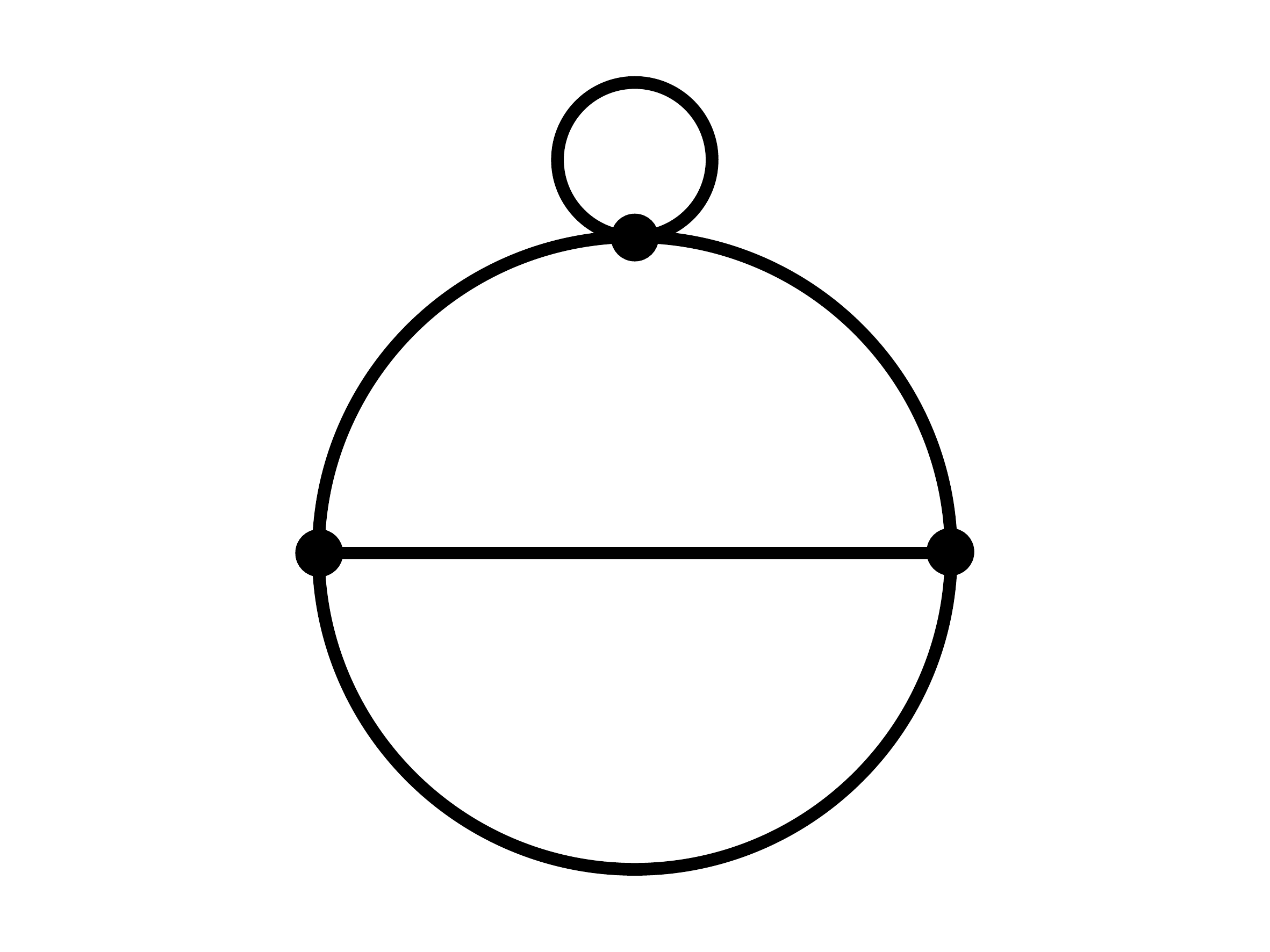}
        \caption{Snowman}
        \label{fig:snowman}
    \end{subfigure}
    \hfill
    \begin{subfigure}[b]{0.4\textwidth}
        \includegraphics[width=\textwidth]{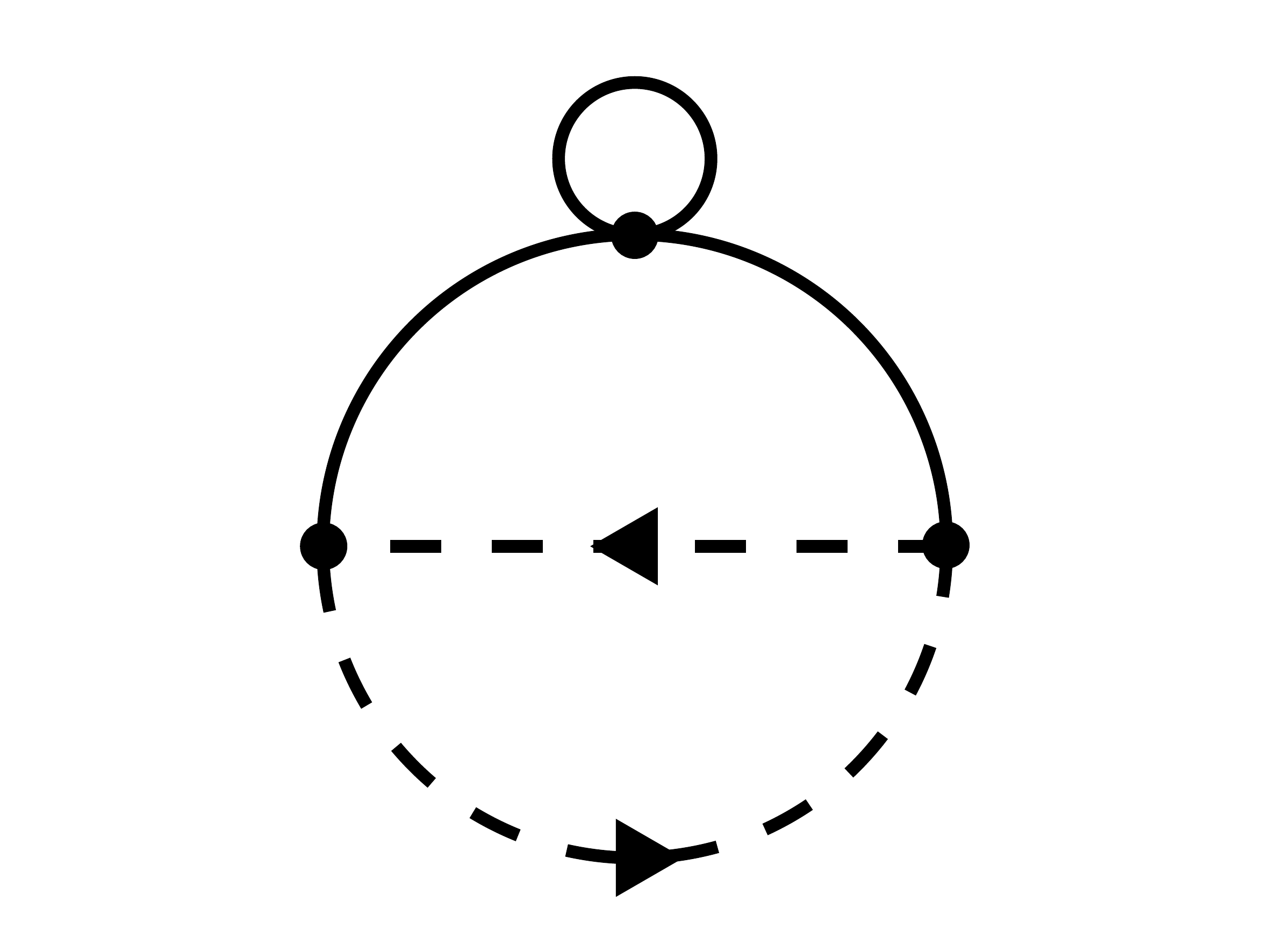}
        \caption{Ghostly snowman}
        \label{fig:snowmanghost}
    \end{subfigure}
    \caption{Snowman-type three-loop diagrams}
\label{fig:threeloopsnowman}
\end{figure}

\begin{figure}
\centering
     \begin{subfigure}[b]{0.4\textwidth}
         \centering
         \includegraphics[width=\textwidth]{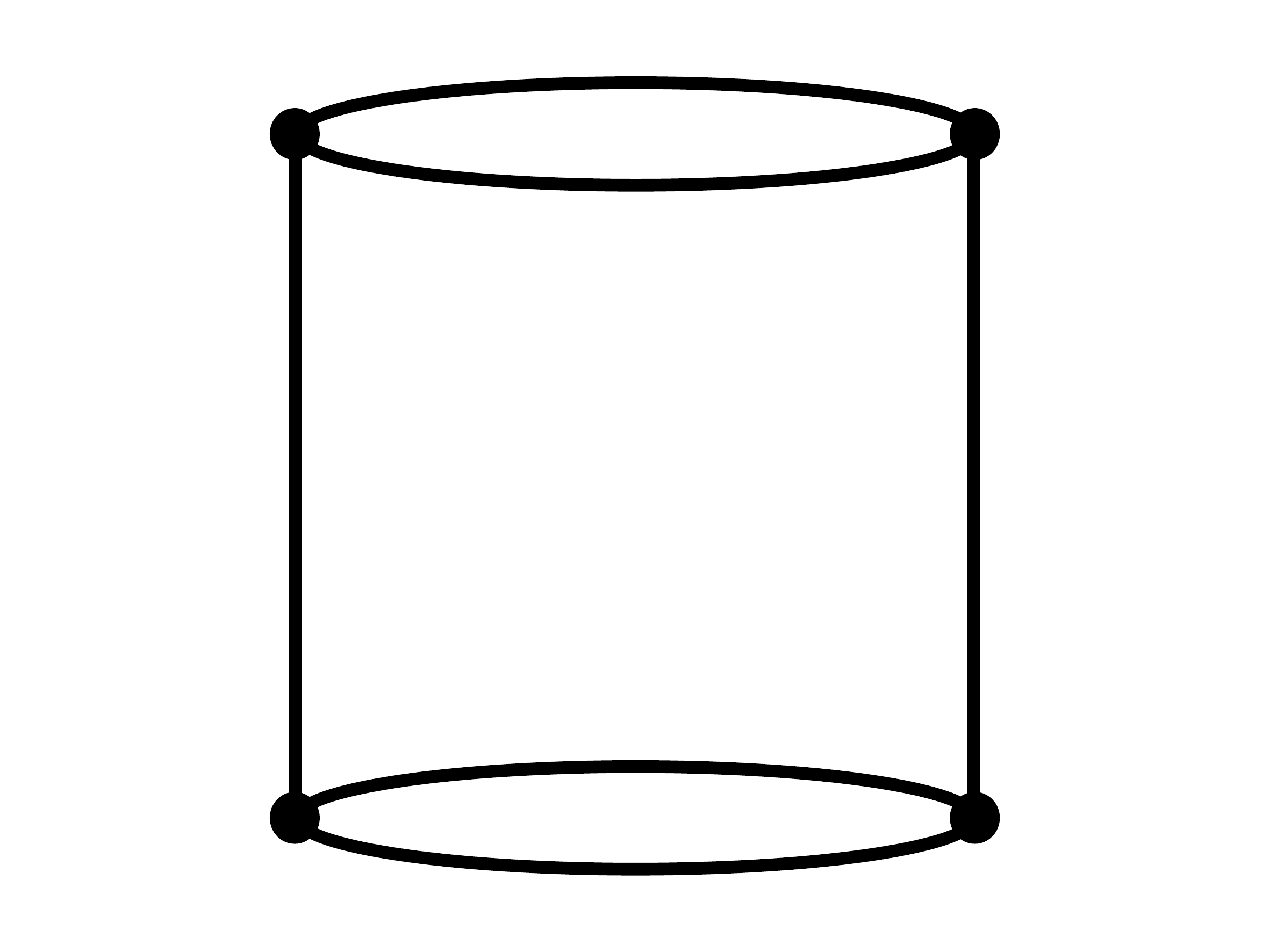}
         \caption{Cylinder}
     \end{subfigure}
     \hfill
     \begin{subfigure}[b]{0.4\textwidth}
         \centering
         \includegraphics[width=\textwidth]{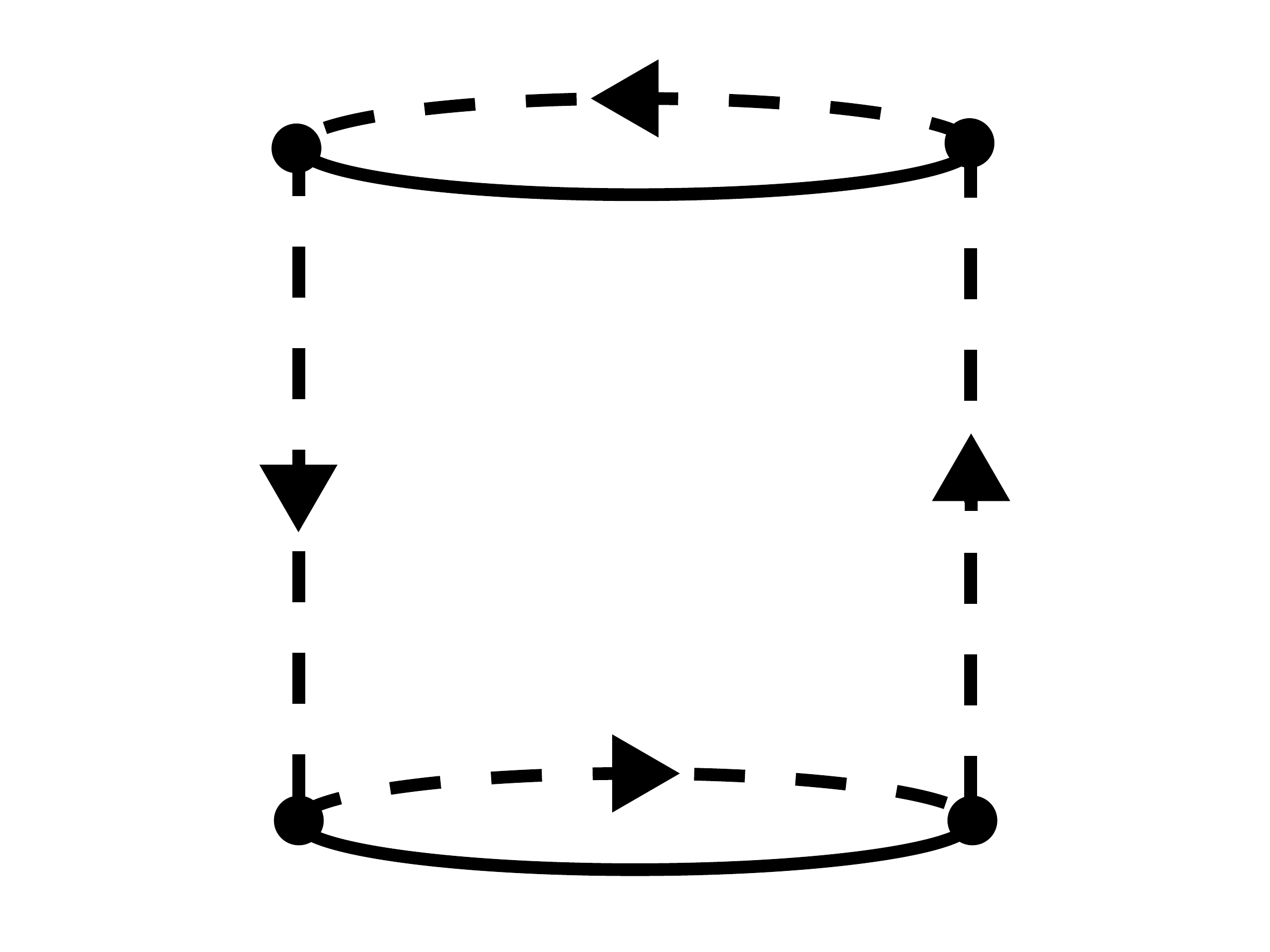}
         \caption{Ghoslty circulating cylinder}
     \end{subfigure}
     \hfill
         \begin{subfigure}[b]{0.4\textwidth}
         \centering
         \includegraphics[width=\textwidth]{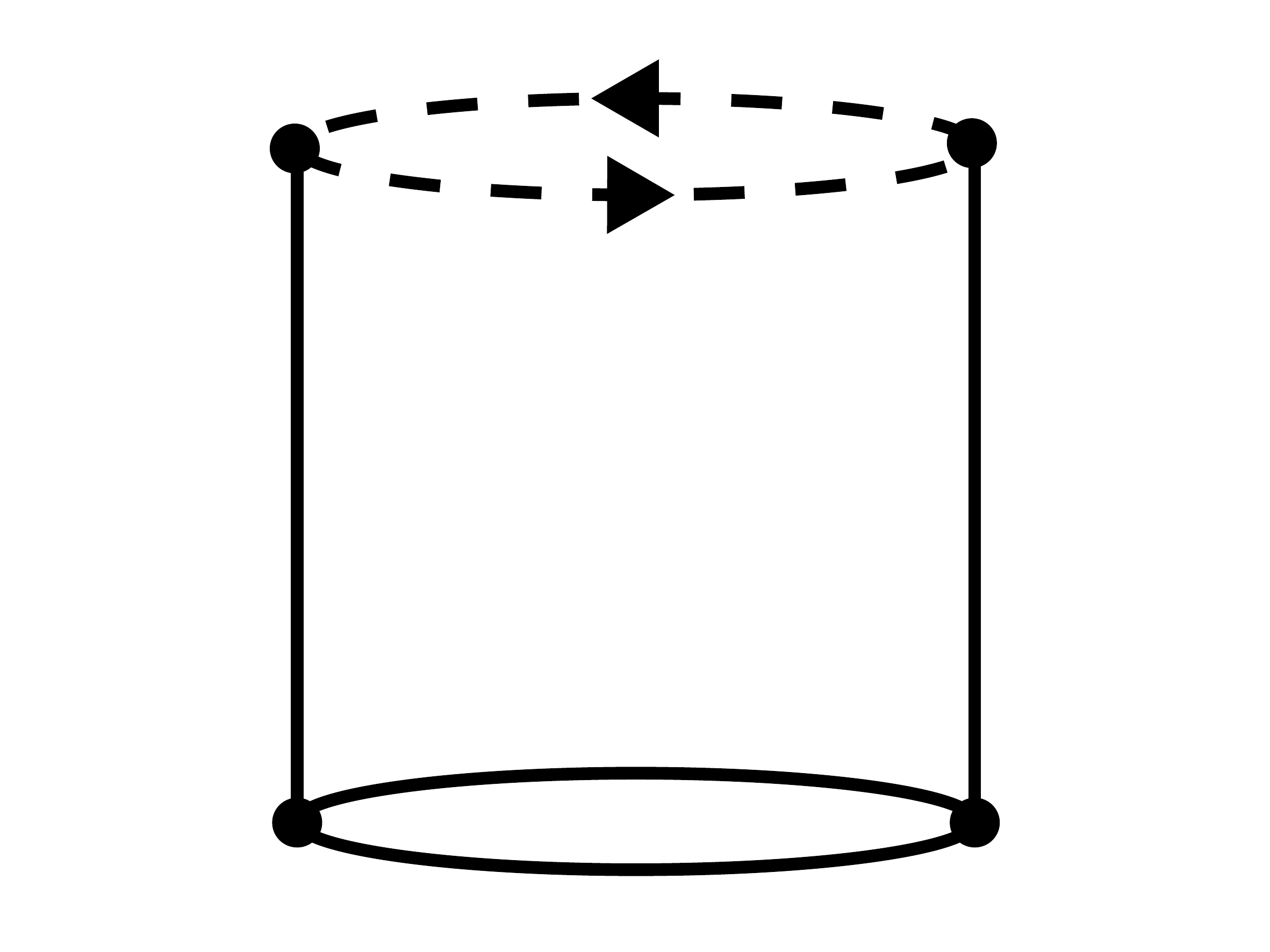}
         \caption{Cylinder with one ghostly lid}
     \end{subfigure}
     \hfill
     \begin{subfigure}[b]{0.4\textwidth}
         \centering
         \includegraphics[width=\textwidth]{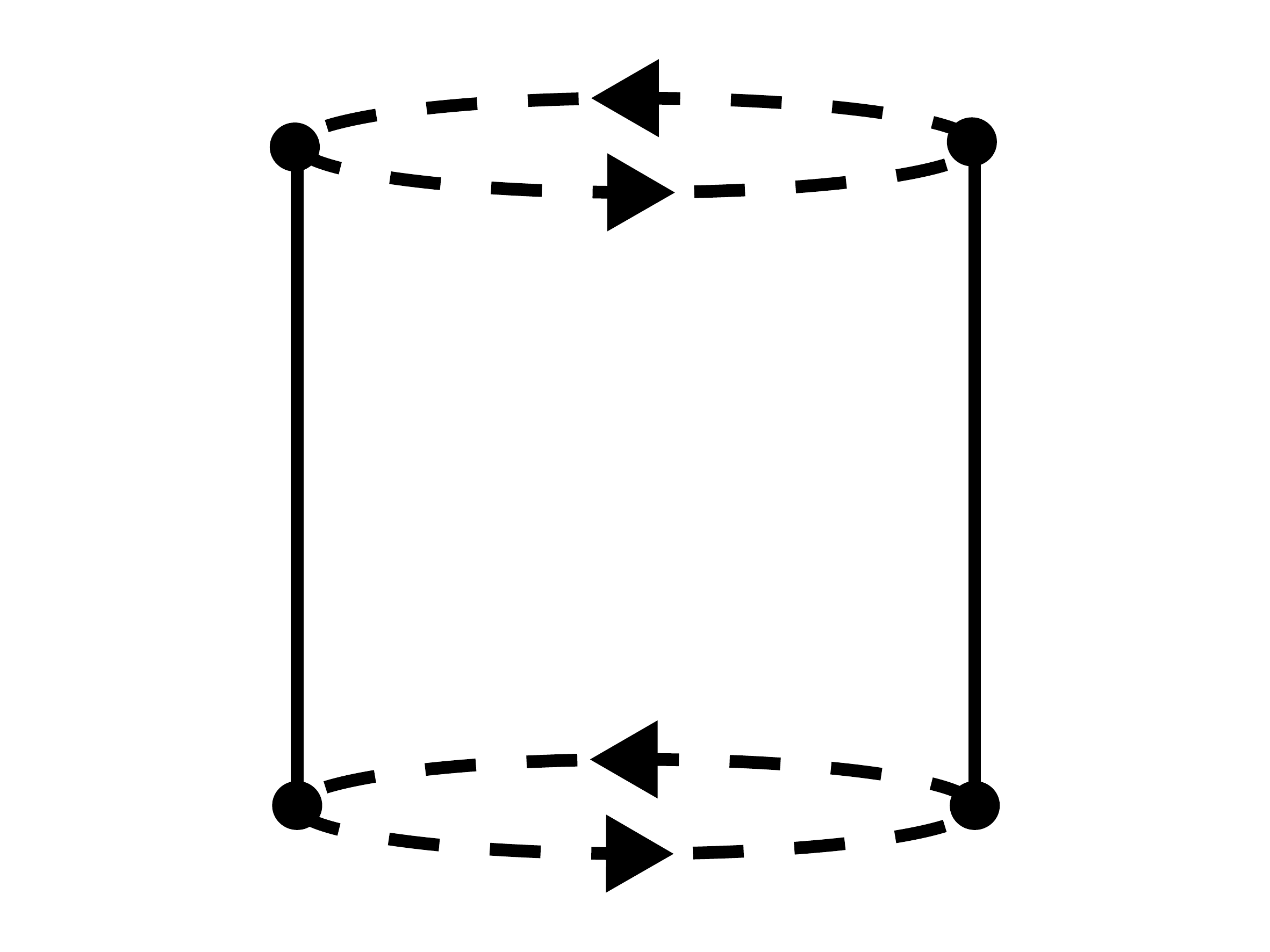}
         \caption{Cylinder with two ghostly lid}
     \end{subfigure}
    \caption{Cylinder-type three-loop diagrams}
    \label{fig:threeloopcylinder}
\end{figure}

\begin{figure}
     \centering
     \begin{subfigure}[b]{0.4\textwidth}
         \centering
         \includegraphics[width=\textwidth]{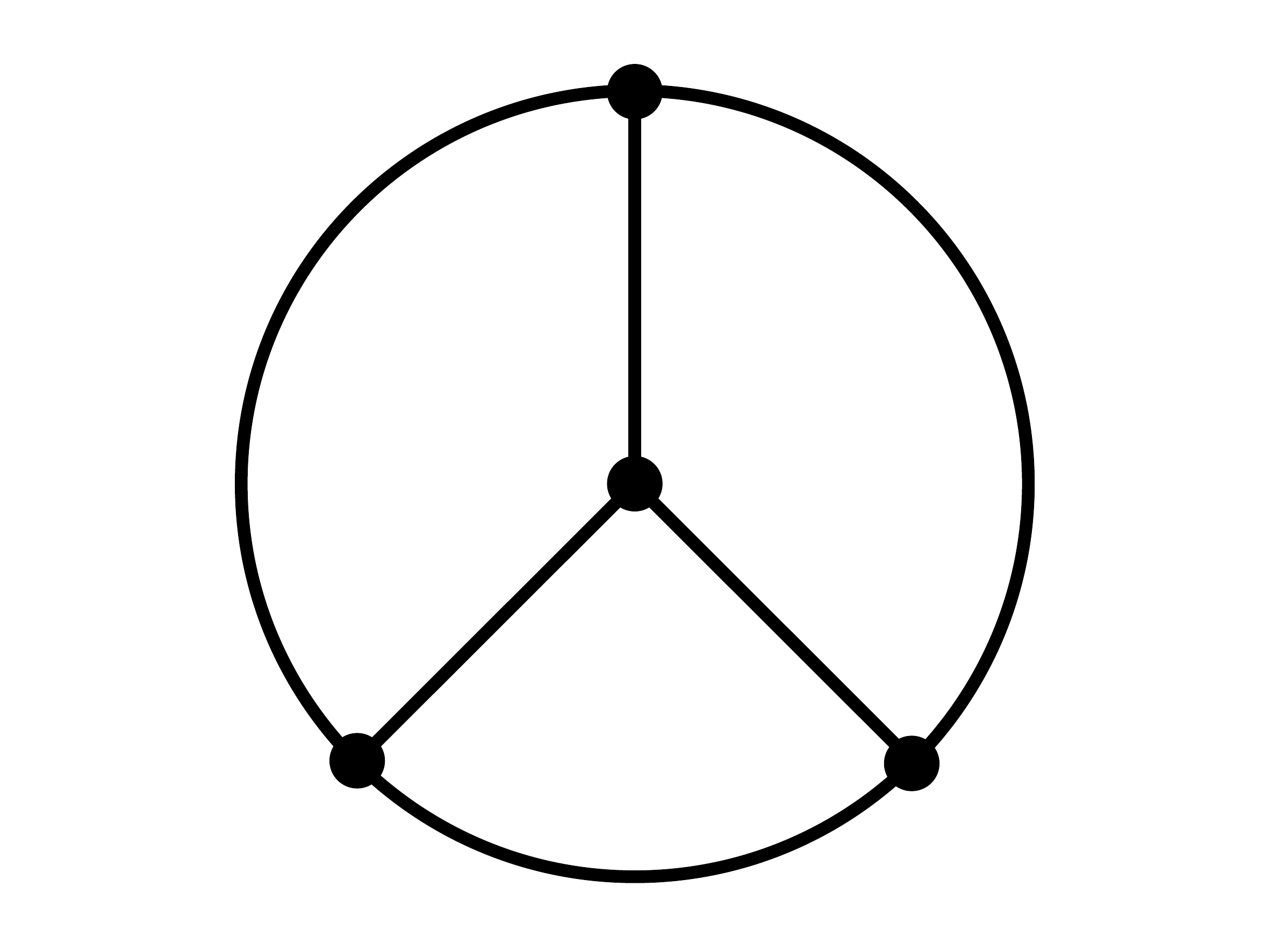}
         \caption{Mercedes Benz diagram}
         \label{fig:MB}
     \end{subfigure}
     \hfill
     \begin{subfigure}[b]{0.4\textwidth}
         \centering
         \includegraphics[width=\textwidth]{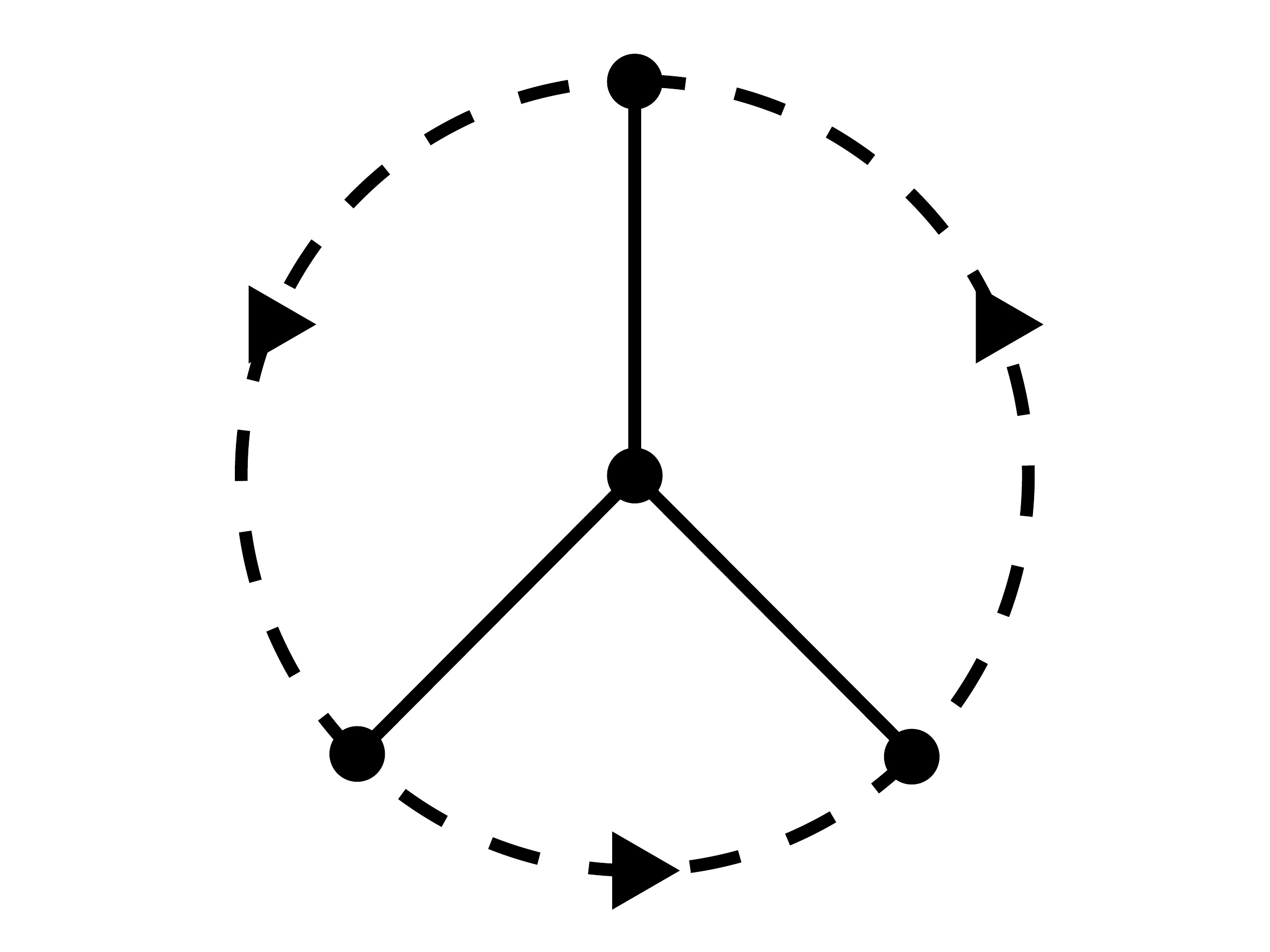}
         \caption{Type 1 ghostly circulating Mercedes Benz diagram}
     \end{subfigure}
     \hfill
     \begin{subfigure}[b]{0.4\textwidth}
         \centering
         \includegraphics[width=\textwidth]{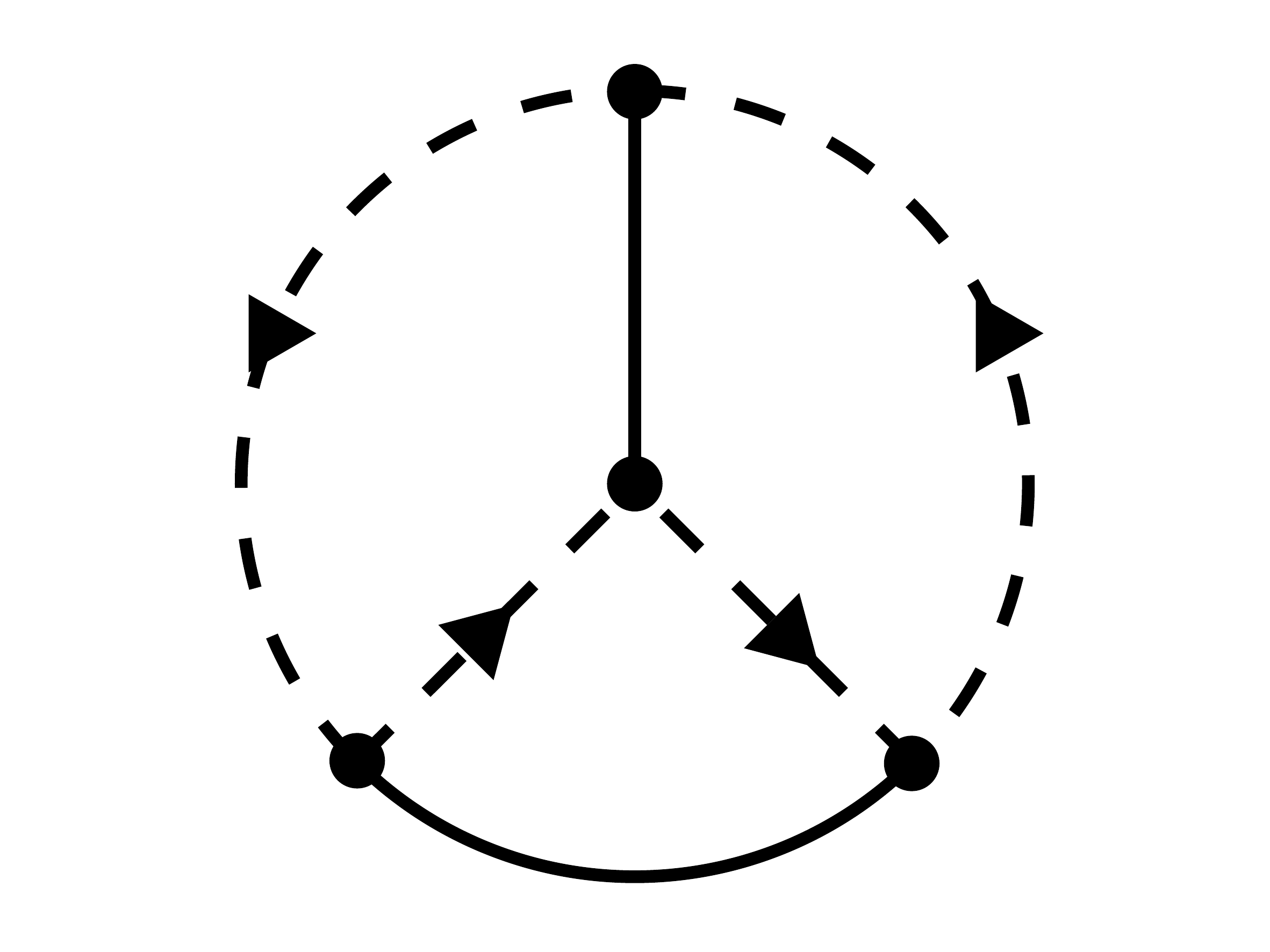}
         \caption{Type 2 ghostly circulating Mercedes Benz diagram}
         \label{fig:five over x}
     \end{subfigure}
        \caption{Mercedes Benz-type three-loop diagram}
        \label{fig:threeloopMB}
\end{figure}

\clearpage

\section{Tables}
\begin{table}[h]
    \centering
        \begin{tabular}{m{2.5cm} P{4cm} P{2.5cm} } 
            \textbf{Diagram} & \textbf{Symmetry factor} & \textbf{Integral} \\ \midrule
          \includegraphics[width=1.65cm]{ThreeLoopsGlasses.pdf}
         & $\dfrac{1}{8}$ & $-\dfrac{61}{2}\, I$ \\ 
            \includegraphics[width=1.65cm]{ThreeLoopsCylinder.pdf} & $\dfrac{1}{16}$   & $16\, I$ \\
            \includegraphics[width=1.65cm]{ThreeLoopsCylinderGhostCirculating.pdf} & $-\dfrac{1}{2}$ & $I$   \\
            \includegraphics[width=1.65cm]{ThreeLoopsCylinderGhost1Lid.pdf} & $-\dfrac{1}{4}$   & $7\, I$   \\
            \includegraphics[width=1.65cm]{ThreeLoopsCylinderGhost2Lids.pdf} & $\dfrac{1}{4}$ & $-\dfrac{3}{2}\, I$ \\ 
            \includegraphics[width=1.65cm]{ThreeLoopsMB.pdf} & $\dfrac{1}{24}$ & $36\, I$ \\
            \includegraphics[width=1.65cm]{ThreeLoopsMBcirculating.pdf} & $-\dfrac{1}{3}$ & $-\dfrac{33}{4}\, I$ \\
            \includegraphics[width=1.65cm]{ThreeLoopsMBcirculating2.pdf} & $-\dfrac{1}{4}$ & $\dfrac{13}{2}\, I$ \\ 
            \includegraphics[width=1.65cm]{ThreeLoopsMellon.pdf} & $\dfrac{1}{48}$ & $135 \, I$
    \end{tabular}
    \caption{Contribution of the non-vanishing 3-loops diagrams in $D=3$. The symmetry factor of each diagram is stated in the second column. The result of the momentum integral is depicted in the third column (in $D=3$ all the integrals are proportional to the master integral $I$).}
    \label{tab:diagrams.contribution.3d}
\end{table}

\clearpage

The contribution of the three-loop diagram in arbitrary dimensions is split into two different tables. The first one contains diagrams proportional to the master integral $I$ while the second one depicts those diagrams whose contributions are not given solely by $I$.

\begin{center}
\begin{table}[h!]
        \centering 
        \begin{tabular}{m{2.5cm} P{3cm} P{7.5cm} } 
            \textbf{Diagram} & \textbf{Symmetry factor} & \textbf{Integral} \\ \midrule
            \includegraphics[width=2cm]{ThreeLoopsCylinderGhostCirculating.pdf} & $-\dfrac{1}{2}$ & $-\frac{(D^5 - 2D^4 - 11D^3 + 24D^2 + 8D - 16)}{16 (D - 2)^2} I$ \\
             \includegraphics[width=2cm]{ThreeLoopsCylinderGhost1Lid.pdf} & $-\dfrac{1}{4}$ & $-\frac{(D^4 - 2D^3 + 3D^2 + 10D - 56)}{16(D - 2)} I$ \\
             \includegraphics[width=2cm]{ThreeLoopsCylinderGhost2Lids.pdf} & $\dfrac{1}{4}$ & $\frac{D^3 - D^2 + 12D - 60}{16(D - 2)} I$ \\
             \includegraphics[width=2cm]{ThreeLoopsMB.pdf} & $\dfrac{1}{24}$ & $-\frac{(D^6 - 69D^5 + 740D^4 - 2710D^3 + 3188D^2 - 64D + 576)}{128(D - 2)^2} I$ \\
             \includegraphics[width=2cm]{ThreeLoopsMBcirculating.pdf} & $-\dfrac{1}{3}$ & $\frac{D^5 - D^4 - 10D^3 - 8D^2 + 80D - 16}{16(D - 2)^2} I$ \\
             \includegraphics[width=2cm]{ThreeLoopsMBcirculating2.pdf} & $-\dfrac{1}{4}$ & $\frac{3D^4 - 35D^3 - 2D^2 + 472D - 800}{64(D - 2)^2} I$ \\
             \includegraphics[width=2cm]{ThreeLoopsMellon.pdf} & $\dfrac{1}{48}$ & $\frac{D^5 + 22D^4 - 169D^3 + 184D^2 + 308D + 48}{32(D - 2)} I$ \\
        \end{tabular}
        \captionof{table}{Non-vanishing diagrams whose momentum integral is proportional the master integral $I$.}
    \label{tab:diagrams.contribution.I}
    \end{table}
\end{center}

\clearpage

\begin{landscape}
    \begin{table}[ht]
        \centering 
        \begin{tabular}{m{2.5cm} P{3cm} P{12.5cm} } 
            \textbf{Diagram} & \textbf{Symmetry factor} & \textbf{Integral} \\ \midrule
            \includegraphics[width=1.75cm]{ThreeLoopsTurtle.pdf} & $\dfrac{1}{12}$ & $\frac{(D - 3) D (9D^3 - 79D^2 + 138D + 112)}{64(D - 2)} J_1$ \\
             \includegraphics[width=2cm]{ThreeLoopsBear.pdf} & $\dfrac{1}{16}$ & $-\frac{(D - 3)^2(D - 5)D^3}{64(D - 2)} K_1$ \\
             \includegraphics[width=2cm]{ThreeLoopsSnowman.pdf} & $\dfrac{1}{8}$ & $-\frac{(D - 3) D (7D^4 - 76D^3 + 253D^2 - 246D - 72)}{64 (D - 2)^2} J_1$ \\
             \includegraphics[width=2cm]{ThreeLoopsSnowmanGhost.pdf} & $-\dfrac{1}{144}$ & $-\frac{(D - 3) D (D^2 - 4D - 1)}{576(D - 2)} J_1 + \frac{(D - 3) D}{576} K_2$ \\
            \includegraphics[width=2cm]{ThreeLoopsGlasses.pdf} & $\dfrac{1}{8}$ & $-\frac{(3D^6 + 108D^5 - 1135D^4 + 3284D^3 - 2256D^2 - 1112D - 1280)}{64(D - 2)^2} I - \frac{(D - 3) D (3 D^3 - 97 D^2 + 270 D + 232)}{64(D - 2)} J_1$ \\
            \includegraphics[width=2cm]{ThreeLoopsCylinder.pdf} & $\dfrac{1}{16}$ & $\frac{5D^6 + 68D^5 - 789D^4 + 2200D^3 - 1228D^2 - 1472D - 64}{128(D - 2)^2} I + \frac{(D - 3)^2 D(3D^3 - 37D^2 + 78D + 16)}{64(D - 2)^2} J_1$
            \\
             & & $+ \frac{(D - 3) D (D^3 - 9 D^2  + 18 D)}{64(D - 2)} K_1 + \frac{(D - 3) D (2 D^3 - 18 D^2 + 28 D + 32)}{64(D - 2)} K_2 $ \\
              & & $+ \frac{(D - 3)^2 D^2}{64} A + \frac{-(D - 3)^2 D^2}{64} C$
        \end{tabular}
        \captionof{table}{Non-vanishing diagrams whose contributions in $D$ dimensions are not given solely by the master integral $I$. The coefficient of the first four diagrams is zero in $D=3$.}
    \label{tab:diagrams.contribution.D}
    \end{table}
\end{landscape}

\clearpage

\section{Master Integrals}
In the following, $\int$ means $\SumInt$:

\begin{subequations}
    \begin{align}
        I &= \int \prod_{i=1}^3d^D k_i \frac{(k_1 \cdot k_2)(k_1 \cdot k_3)}{k_1^2k_2^2k_3^2(k_1+k_2+k_3)^2} \\
        J_1 &= \int\prod_{i=1}^3d^D k_i\frac{(k_1 \cdot k_3)^2}{k_1^2k_2^2(k_1+k_2)^2k_3^2}\\
        K_1 &= \int\prod_{i=1}^3d^D k_i \frac{(k_2 \cdot k_3)^2}{k_1^4 k_2^2 k_3^2}\\
        K_2 &= \int\prod_{i=1}^3d^D k_i \frac{(k_1 \cdot k_2)^2}{k_1^4 k_2^2 k_3^2}\\
        A &= \int \prod_{i=1}^3d^D k_i \frac{(k_2 \cdot k_3)^3}{k_1^2 (k_1 + k_2)^2 (k_1 + k_3)^2 k_2^2 k_3^2}\\
        B &= \int \prod_{i=1}^3d^D k_i \frac{k_1^6}{(k_1 + k_2)^2 (k_1 + k_3)^2 k_2^2 k_3^2 (k_2 - k_3)^2}\\
        C &= \int \prod_{i=1}^3d^D k_i \frac{(k_2 \cdot k_3)^4} {k_1^4 (k_1+k_2)^2 (k_1+k_3)^2 k_2^2 k_3^2} \,.
     \end{align}
     \label{eq:master.integrals}
\end{subequations}

\clearpage

\providecommand{\href}[2]{#2}\begingroup\raggedright\endgroup


\begin{thebibliography}{10}

\bibitem{Maloney.Witten}
A.~Maloney and E.~Witten, \emph{{Quantum Gravity Partition Functions in Three Dimensions}}, JHEP {\bf 02} (2010) 029,
\href{http://www.arXiv.org/abs/0712.0155}{{\tt 0712.0155}}

\bibitem{Brown.Henneaux}
J.~D. Brown and M.~Henneaux, \emph{{Central Charges in the Canonical Realization of Asymptotic Symmetries: An Example from Three-Dimensional Gravity}}, Commun. Math. Phys. {\bf 104} (1986)
207--226

\bibitem{Cotler.Jensen:AdS3gravity.randomCFT}
J.~Cotler and K.~Jensen, \emph{AdS3 gravity and random CFT}, Journal of High Energy Physics {\bf 2021}
(Apr., 2021)

\bibitem{MaloneyAverage}
J.~Chandra, S.~Collier, T.~Hartman  and A.~Maloney, \emph{{Semiclassical 3D gravity as an average of large-c CFTs}}, JHEP {\bf 12} (2022) 069,
\href{http://www.arXiv.org/abs/2203.06511}{{\tt 2203.06511}}

\bibitem{MaloneyConical}
N.~Benjamin, S.~Collier  and A.~Maloney, \emph{{Pure Gravity and Conical Defects}}, JHEP {\bf 09} (2020) 034,
\href{http://www.arXiv.org/abs/2004.14428}{{\tt 2004.14428}}

\bibitem{Witten:TopologyChaningAmplitudes}
E.~Witten, \emph{{Topology Changing Amplitudes in (2+1)-Dimensional Gravity}}, Nucl. Phys. B {\bf 323} (1989)
113--140

\bibitem{Barnich:One.Loop.Flat}
G.~Barnich, H.~A. Gonzalez, A.~Maloney  and B.~Oblak, \emph{{One-loop partition function of three-dimensional flat gravity}}, JHEP {\bf 04} (2015) 178,
\href{http://www.arXiv.org/abs/1502.06185}{{\tt 1502.06185}}

\bibitem{Leston:ThreeLoops}
M.~Leston, A.~Goya, G.~P\'erez-Nadal, M.~Passaglia  and G.~Giribet, \emph{{3D Quantum Gravity Partition Function at Three Loops}}, Phys. Rev. Lett. {\bf 131} (2023), no.~18, 181601,
\href{http://www.arXiv.org/abs/2307.03830}{{\tt 2307.03830}}

\bibitem{Buchbinder:TwoLoopApproach}
I.~L. Buchbinder, S.~D. Odintsov  and O.~A. Fonarev, \emph{{Two loop approach to effective action in quantum gravity}}, Sov. J. Nucl. Phys. {\bf 52} (1990)
1101--1106

\bibitem{Buchbinder:TwoLoopsApproximation}
I.~L. Buchbinder, S.~D. Odintsov  and O.~A. Fonarev, \emph{{Two loop approximation in quantum gravitation}}, Phys. Lett. B {\bf 245} (1990)
365--369

\bibitem{Balasubramanian.Kraus}
V.~Balasubramanian and P.~Kraus, \emph{{A Stress tensor for Anti-de Sitter gravity}}, Commun. Math. Phys. {\bf 208} (1999) 413--428,
\href{http://www.arXiv.org/abs/hep-th/9902121}{{\tt hep-th/9902121}}

\bibitem{HolographicRenormalization}
S.~de~Haro, S.~N. Solodukhin  and K.~Skenderis, \emph{{Holographic reconstruction of space-time and renormalization in the AdS / CFT correspondence}}, Commun. Math. Phys. {\bf 217} (2001) 595--622,
\href{http://www.arXiv.org/abs/hep-th/0002230}{{\tt hep-th/0002230}}

\bibitem{Cotler.Jensen:reparametrization.AdS3}
J.~Cotler and K.~Jensen, \emph{A theory of reparameterizations for AdS3 gravity}, Journal of High Energy Physics {\bf 2019}
(Feb., 2019)

\bibitem{Giombi:One.Loop}
S.~Giombi, A.~Maloney  and X.~Yin, \emph{{One-loop Partition Functions of 3D Gravity}}, JHEP {\bf 08} (2008) 007,
\href{http://www.arXiv.org/abs/0804.1773}{{\tt 0804.1773}}

\bibitem{Acosta:SquareIntegrability}
J.~Acosta, A.~Garbarz, A.~Goya  and M.~Leston, \emph{{Asymptotic Boundary Conditions and Square Integrability in the Partition Function of AdS Gravity}}, JHEP {\bf 06} (2020) 172,
\href{http://www.arXiv.org/abs/2004.01723}{{\tt 2004.01723}}

\bibitem{Acosta:Accesibility}
J.~Acosta, A.~Garbarz, A.~Goya  and M.~Leston, \emph{{One-loop partition function, gauge accessibility and spectra in AdS$_{3}$ gravity}}, JHEP {\bf 12} (2021) 097,
\href{http://www.arXiv.org/abs/2109.06938}{{\tt 2109.06938}}

\bibitem{Ashtekar:BMS}
A.~Ashtekar, J.~Bicak  and B.~G. Schmidt, \emph{{Asymptotic structure of symmetry reduced general relativity}}, Phys. Rev. D {\bf 55} (1997) 669--686,
\href{http://www.arXiv.org/abs/gr-qc/9608042}{{\tt gr-qc/9608042}}

\bibitem{Barnich:BMS}
G.~Barnich and G.~Compere, \emph{{Classical central extension for asymptotic symmetries at null infinity in three spacetime dimensions}}, Class. Quant. Grav. {\bf 24} (2007) F15--F23,
\href{http://www.arXiv.org/abs/gr-qc/0610130}{{\tt gr-qc/0610130}}

\bibitem{Cotler.etal:soft.gravitons.3d}
J.~Cotler, K.~Jensen, S.~Prohazka, M.~Riegler  and J.~Salzer, \emph{{Soft gravitons in three dimensions}}, JHEP {\bf 07} (2025) 002,
\href{http://www.arXiv.org/abs/2411.13633}{{\tt 2411.13633}}

\bibitem{Landau.Lifschits:ClassicalFields}
L.~D. Landau and E.~M. Lifschits, {\em {The Classical Theory of Fields}}, vol.~Volume 2 of {\em Course of Theoretical Physics}.
\newblock Pergamon Press, Oxford,
1975

\bibitem{FeynCalc1}
R.~Mertig, M.~Bohm  and A.~Denner, \emph{{FEYN CALC: Computer algebraic calculation of Feynman amplitudes}}, Comput. Phys. Commun. {\bf 64} (1991)
345--359

\bibitem{FeynCalc2}
V.~Shtabovenko, R.~Mertig  and F.~Orellana, \emph{{New Developments in FeynCalc 9.0}}, Comput. Phys. Commun. {\bf 207} (2016) 432--444,
\href{http://www.arXiv.org/abs/1601.01167}{{\tt 1601.01167}}

\bibitem{FeynCalc3}
V.~Shtabovenko, R.~Mertig  and F.~Orellana, \emph{{FeynCalc 9.3: New features and improvements}}, Comput. Phys. Commun. {\bf 256} (2020) 107478,
\href{http://www.arXiv.org/abs/2001.04407}{{\tt 2001.04407}}

\end{thebibliography}
\end{document}